%% file: main.tex
\documentclass[sigconf,natbib=true,anonymous=false]{acmart}
\usepackage{subcaption} 
\usepackage{graphicx} 
\usepackage{xcolor}
\newcommand{\redcomment}[1]{\textcolor{red}{(#1)}}
\usepackage[colorinlistoftodos,prependcaption,textsize=tiny]{todonotes}
\usepackage{cleveref}
\usepackage{bm}
\usepackage{multirow}
\usepackage[acronym]{glossaries}
\makeglossaries
\input{glossary}
\usepackage{booktabs}
\usepackage{adjustbox}
\usepackage{soul}
\usepackage{amsmath}

\usepackage{microtype}
\usepackage{balance}
\usepackage{tabularx}
\usepackage{colortbl}
\usepackage{paralist}
\usepackage{array}
\usepackage{tcolorbox}
\usepackage[inline]{enumitem}
\usepackage{siunitx}

\newcolumntype{L}[1]{>{\raggedright\let\newline\\\arraybackslash\hspace{0pt}}m{#1}}
\newcolumntype{C}[1]{>{\centering\let\newline\\\arraybackslash\hspace{0pt}}m{#1}}
\newcolumntype{R}[1]{>{\raggedleft\let\newline\\\arraybackslash\hspace{0pt}}m{#1}}

\AtBeginDocument{%
  \providecommand\BibTeX{{%
    \normalfont B\kern-0.5em{\scshape i\kern-0.25em b}\kern-0.8em\TeX}}}

\setcopyright{acmlicensed}
\copyrightyear{2018}
\acmYear{2018}
\acmDOI{XXXXXXX.XXXXXXX}

\acmConference[Conference acronym 'XX]{Make sure to enter the correct
  conference title from your rights confirmation emai}{June 03--05,
  2018}{Woodstock, NY}
\acmISBN{978-1-4503-XXXX-X/18/06}

\begin{document}

\title{Beyond One-Size-Fits-All: A Study of Neural and Behavioural Variability Across Different Recommendation Categories}


\author{Georgios Koutroumpas}
\affiliation{%
  \institution{Telef\'{o}nica Scientific Research}
  \city{Barcelona}
  \country{Spain}
}
\email{georgios.koutroumpas@upc.edu}

\author{Sebastian Idesis}
\affiliation{%
  \institution{Telef\'{o}nica Scientific Research}
  \city{Barcelona}
  \country{Spain}
}
\email{sebastianariel.idesis@telefonica.com}

\author{Mireia Masias Bruns}
\affiliation{%
  \institution{Telef\'{o}nica Scientific Research}
  \city{Barcelona}
  \country{Spain}
}
\email{mireia.masiasbruns@telefonica.com}

\author{Carlos Segura}
\affiliation{%
  \institution{Telef\'{o}nica Scientific Research}
  \city{Barcelona}
  \country{Spain}
}
\email{carlos.segura@telefonica.com}

\author{Joemon M. Jose}
\affiliation{%
  \institution{University of Glasgow}
  \city{Glasgow}
  \country{United Kingdom}
}
\email{joemon.jose@glasgow.ac.uk}

\author{Sergi Abadal}
\affiliation{%
  \institution{Universitat Politècnica de Catalunya}
  \city{Barcelona}
  \country{Spain}
}
\email{abadal@ac.upc.edu}

\author{Ioannis Arapakis}
\affiliation{%
  \institution{Telef\'{o}nica Scientific Research}
  \city{Barcelona}
  \country{Spain}
}
\email{ioannis.arapakis@telefonica.com}

\renewcommand{\shortauthors}{Trovato and Tobin, et al.}

\begin{abstract} 
Traditionally, Recommender Systems (RS) have primarily measured performance based on the accuracy and relevance of their recommendations. However, this algorithmic-centric approach overlooks how different types of recommendations impact user engagement and shape the overall quality of experience. In this paper, we shift the focus to the user and address for the first time the challenge of decoding the neural and behavioural variability across \textit{distinct} recommendation categories, considering more than just relevance. Specifically, we conducted a controlled study using a comprehensive e-commerce dataset containing various recommendation types, and collected Electroencephalography and behavioural data. We analysed both neural and behavioural responses to recommendations that were categorised as \textbf{E}xact, \textbf{S}ubstitute, \textbf{C}omplement, or \textbf{I}rrelevant products within search query results. Our findings offer novel insights into user preferences and decision-making processes, revealing meaningful relationships between behavioural and neural patterns for each category, but also indicate inter-subject variability.
\end{abstract}





\maketitle

\section{Introduction}
\label{sec:introduction}

The rapid growth of e-commerce has fundamentally changed how users discover and interact with products and services online. While this digital transformation has created more revenue opportunities for service providers, it has also introduced significant challenges for users who must navigate an overwhelming number of choices. Specifically, when presented with large array situations (e.g., limitless products to purchase from), users are at higher risk of experiencing choice overload \cite{iyengar1999rethinking}, which can degrade their quality of experience with an online service or platform.

\gls{rs} have been shown to alleviate this ``paradox of choice'' \cite{schwartz2004paradox} by facilitating access to relevant content and improving the browsing experience \cite{hu2018reinforcement, yuan2020}. In settings where the abundance of options can result in unsatisfying choices or abandonment, the user experience is ultimately determined by the \gls{rs} capacity to filter irrelevant content and recommend items regarded as desirable. While significant advances have been made in improving recommendation accuracy \cite{98,93.1,deldjoo2024reviewmodernrecommendersystems}, these approaches often optimise mainstream metrics at the expense of other content-derived qualitative aspects, such as diversity and novelty of recommendations \cite{Anderson2020,93.1,93.2,94,98}. More importantly, current approaches primarily focus on algorithmic performance, potentially overlooking the complex cognitive and behavioural responses that different recommendation categories may elicit from users. 

Previous research has explored neural correlates of relevance detection \cite{Tukka, Pinkosova2023} and search satisfaction \cite{Pinkosova2023, Moshfeghi21}, mainly in binary relevance tasks involving search terms and documents. In contrast, we explore how users process  and respond to \textit{different} recommendation categories, by leveraging a real-life, e-commerce query-product dataset (``ESCI'' \cite{Reddy2022}), where the relationship between user intent and recommended products is more nuanced than simple relevance. Our research addresses the following questions:

\begin{itemize}
    \item \textbf{RQ1:} Do different recommendation categories evoke differentiable neural signatures in brain activity as measured through \gls{eeg}, and can these patterns be reliably identified?
       
    \item \textbf{RQ2:} How do different recommendation categories influence users' perception (and subsequent observable behaviour) regarding item relevance, purchaseability, and recommendation diversity?
  
    \item \textbf{RQ3:} Can we identify neurophysiological markers of engagement (e.g., \gls{faa} \cite{barros2022frontal}) that systematically vary across recommendation categories, and what do these variations reveal about the effectiveness of different recommendation strategies?
\end{itemize}

Specifically, by combining \gls{eeg} measurements with behavioural data, our study offers a fresh perspective on recommendation effectiveness that goes beyond traditional accuracy metrics. This approach allows us to examine whether different recommendation categories evoke distinct neural signatures, how they influence user perception of relevance and purchase intent, and their impact on overall engagement. Our findings could help inform the development of more sophisticated \gls{rs} in the future so that they better align with users' intent and true consumer needs. 

In summary, our \textbf{contributions} are the following:

\begin{itemize}
    \item We present the first comprehensive study examining neural and behavioural responses across different recommendation categories, leveraging a real-life, e-commerce dataset, and reveal distinct patterns of brain activity for \textbf{E}xact, \textbf{S}ubstitute, \textbf{C}omplement, or \textbf{I}rrelevant product recommendations.
    \item We introduce novel methodological approaches for analysing the relationship between recommendation types and user cognitive processes, combining \gls{eeg} measurements with behavioural metrics in a controlled experimental setting.
    \item We provide empirical evidence demonstrating how different recommendation categories influence user perception of relevance (and subsequent observable behaviour), with implications for balancing accuracy and diversity in \gls{rs}.
    \item We analyse neural markers of engagement (e.g. \gls{faa}, increased theta/beta power) and their association to different recommendation types.
\end{itemize}

\section{Related Work}
\label{sec:related_work}

Despite progress, key challenges remain in \gls{rs}, including the cold-start problem \cite{93.1,93.2,98}, bias in sparse ratings \cite{93.1,95,98}, and limitations of accuracy-based evaluation \cite{93.1,94,98}. While \gls{dl} and \gls{llm} offer potential \cite{deldjoo2024reviewmodernrecommendersystems}, they require significant resources and still struggle with concepts like diversity and novelty. Understanding user cognition and integrating neural signals into \gls{rs} is a promising direction \cite{RS_16,3,14,4,Pinkosova2023}, but little research has explored how different recommendation types shape neural and behavioural responses. In what follows, we review explicit, implicit, and neuroscience-informed feedback methods for relevance prediction.

\subsection{Explicit Feedback Methods}
\label{ssec:explicit_feedback_methods}

Explicit feedback involves direct user input, such as ratings, reviews, or preference selections. While widely studied in \gls{rs} research, it has key limitations \cite{Oard1998ImplicitFF}. A main challenge is the need for active user participation, which many users avoid \cite{Hu2008}, resulting in data sparsity that hampers recommender system performance \cite{adomavicius}. Moreover, explicit feedback often reflects biases and reliability issues, as users may struggle to express preferences, especially for unconscious or hard-to-quantify experiences \cite{aldayel2020deep, pereda2015human}. These drawbacks have motivated the investigation of alternative methods for capturing user preferences \cite{lerche2016using}.

\subsection{Implicit Feedback Methods}
\label{ssec:implicit_feedback_methods}

Implicit feedback refers to user interactions passively collected without direct input, offering an alternative to explicit ratings in \gls{rs} \cite{jawaheer2014modeling}. Such signals include clicks, dwell time, scrolling, and purchase history \cite{white2002comparing}. Its main advantage is non-intrusiveness, reducing user effort and cognitive load \cite{jawaheer2010comparison}. However, these signals can be ambiguous; interactions may not reliably indicate true interest, introducing noise \cite{lerche2016using}. For example, long dwell time could reflect either engagement or confusion \cite{claypool2001implicit}. To overcome these issues, researchers are investigating neurophysiological measures like \gls{eeg} to directly assess cognitive states and engagement.

\subsection{Neuroimaging Methods for Relevance Prediction}
\label{ssec:neuroimaging_methods_for_relevance_prediction}

Traditional \gls{rs} often overlook the subjective emotional and cognitive responses that are crucial for understanding user engagement and satisfaction \cite{3}. Recent studies have leveraged \gls{eeg} to investigate the neural underpinnings of relevance judgments. For example, \citet{Pinkosova2023} recorded \gls{eeg} data during a binary relevance assessment task, demonstrating that neural signals can provide valuable insights into how users perceive relevant content. Similarly, \citet{ye2022don} compared brain signals of users examining non-clicked search results of varying usefulness, providing insights into the neural correlates of relevance judgments. 

Expanding on this work, \citet{Jacucci2019} developed a fully integrated \gls{ir} system that utilizes online implicit relevance feedback based on brain activity and eye movements, enabling real-time adaptation of search results. In a similar vein, \citet{tukka2023} proposed decoding emotional responses from \gls{eeg} as an alternative dimension of relevance, highlighting the role of affect in user preference modelling. More recently, \citet{ye2024brain} reviewed the application of \gls{bci} in \gls{ir} research, identifying several opportunities for integrating active or passive \glspl{bci} into \gls{rs} to provide real-time insights into user cognitive states, thereby improving the relevance prediction. 

These studies demonstrate that neuroimaging techniques, particularly \gls{eeg}, offer promising new avenues for relevance prediction by directly tapping into user cognition, emotion, and engagement. Integrating neurophysiological signals into \gls{rs} models could enable systems to move beyond conventional behavioural tracking toward a more comprehensive understanding of the user experience and inform the design of \gls{rs} algorithms, particularly those incorporating multi-objective criteria.

\section{Methodology}
\label{sec:Methodology}

Our experimental methodology is based on the acquisition of \gls{eeg}
signals from participants as they evaluate 
query-product pairs. Next, we describe the neurophysiological experiment, including participant recruitment and experimental design. Then, we outline the data curation and pre-processing steps
. Finally, we detail the signal processing techniques and methods used to extract relevant neural and behavioural features in response to different recommendations.

\subsection{Design}
\label{ssec:design}

The study used a within-subjects design with one independent variables: (1) recommendation category (with four levels: \textbf{E}xact, \textbf{S}ubstitute, \textbf{C}omplement, or \textbf{I}rrelevant). 
The dependent variables were (1) user neural activity, as registered by the \gls{eeg} and (2) user behaviour, as measured via self-reports.

\subsection{Participants}
\label{ssec:participants}

Twenty-one participants enrolled in this study (8 females; 3 left-handed), aged 18 to 59 years ($M=28.02$, $SD=8.62$) and of mixed nationality. The participants were recruited via [Redacted]'s Testers platform and mailing lists, and provided written consent prior to the experiment. All participants had normal or corrected-to-normal vision, did not suffer from any type of disability, nor sensory or neurological difficulty. Moreover, participants were not under any kind of neurological medication or treatment that could interfere with the study (e.g., antiepileptics
). Participants self-reported at least a C1 level of English proficiency and prior experience with e-commerce platforms (e.g., Amazon). Upon completion of the study, they were compensated with 40\,EUR in Amazon vouchers.

\subsection{Apparatus}
\label{ssec:apparatus}

\subsubsection{Equipment}
\label{sssec:equipment}

For the study, stimuli were presented on a 27-inch monitor (23.53 $\times$ 13.24 inches; 1920 $\times$ 1080 pixels; 60 Hz), positioned ~~60 cm from participants. Responses were collected via a standard mouse and keyboard. Stimulus delivery, hardware synchronisation, and timing optimisation were controlled using PsychoPy\footnote{\url{https://www.psychopy.org/}}\cite{peirce2019psychopy2}). \gls{eeg} data were recorded with an actiCHamp Plus system (Brain Products GmbH, Germany) using 32 active Ag/AgCl electrodes arranged according to the 10/20 system. Signals were sampled at 256 Hz, referenced online to Cz, and grounded at AFz. Electrode impedances were kept below~~\qty{10}{\kilo\ohm} and monitored via BrainVision Recorder\footnote{\url{https://www.brainproducts.com/}} (Brain Products GmbH, Germany) to ensure reliable \gls{snr}. Precise synchronization of stimulus events and \gls{eeg} data was achieved by transmitting millisecond-accurate hardware triggers from PsychoPy to BrainVision Recorder via a parallel port.

\subsubsection{Dataset}
\label{sssec:dataset}

We used the Amazon dataset \cite{ni-etal-2019-justifying}, a widely adopted real-world e-commerce corpus in \gls{rs} research \cite{10545583, 10.1145/3580488, Li2021PersonalizedTF, 10.1145/3580305.3599535, 10.1145/3404835.3462939, Antognini2020TRECSAT, ijcai2021p72}, containing product details such as image URLs, titles, prices, and categories. Additionally, we incorporated the Shopping Queries dataset \cite{Reddy2022}, comprising 130k unique queries and 2.6 million manually labeled query-product relevance judgments. Each query lists up to 40 results with ESCI labels (\textbf{E}xact, \textbf{S}ubstitute, \textbf{C}omplement, or \textbf{I}rrelevant) \cite{mcauley2015}, along with product and query metadata (IDs, titles, descriptions, texts, labels). This multilingual dataset includes English, Japanese, and Spanish queries. After joining both datasets via product IDs and removing inconsistencies or incomplete entries, we excluded non-English queries with fewer than two \textbf{E}xact products. The final dataset (\Cref{tab:esci_datasets}) included product identifiers, titles, descriptions, prices, images, queries, and recommendation types.

\fboxsep=0pt
\fboxrule=0.25pt
\begin{figure*}[t!]
    \def\w{0.184\linewidth}
    \subfloat[Query]{
         \fbox{\includegraphics[trim={320 205 320 205}, clip, width=\w]{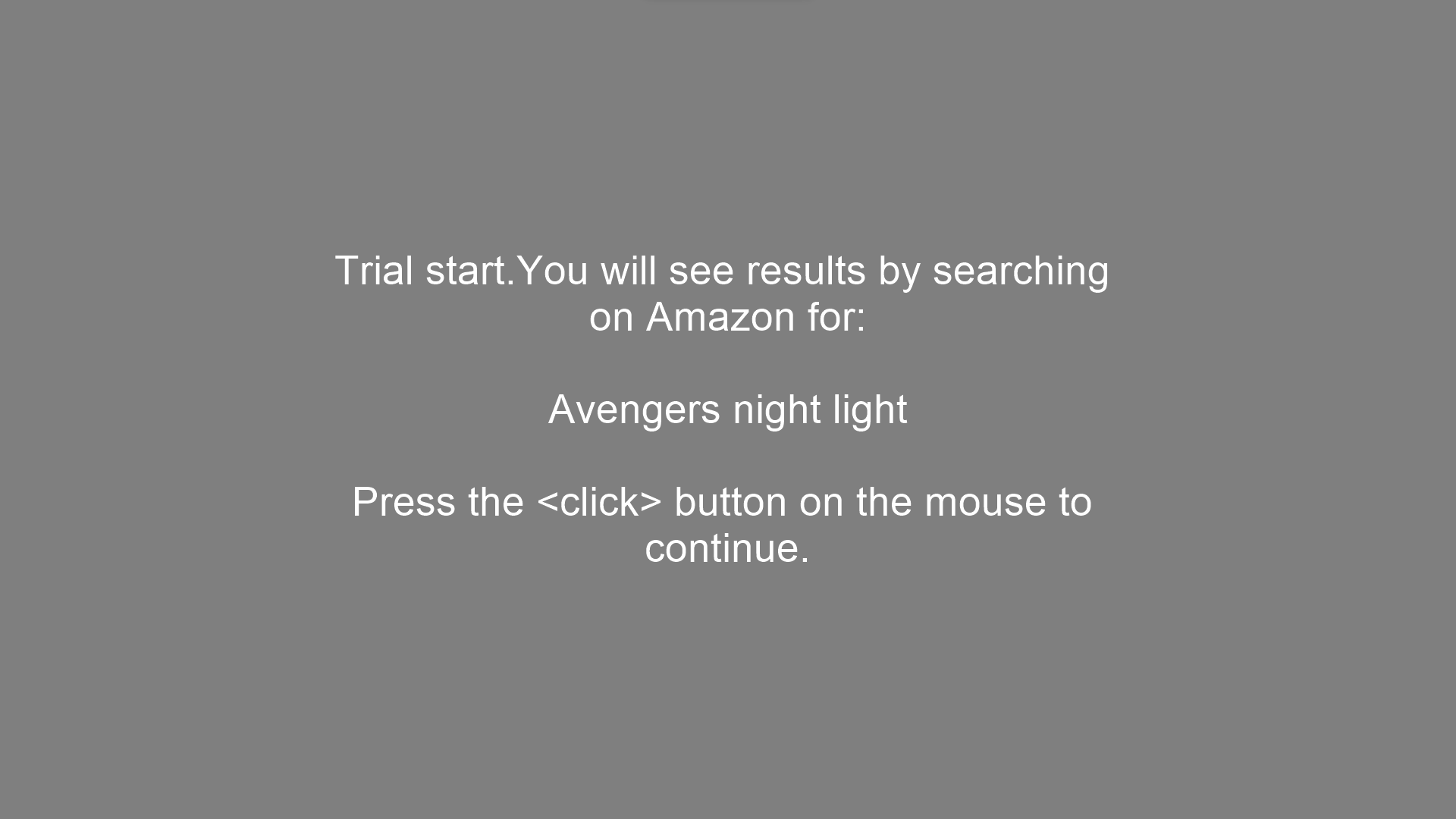}}
    }
    \hspace{0.0025\linewidth}
    \subfloat[Exact Product]{
         \fbox{\includegraphics[trim={320 205 320 205}, clip, width=\w]{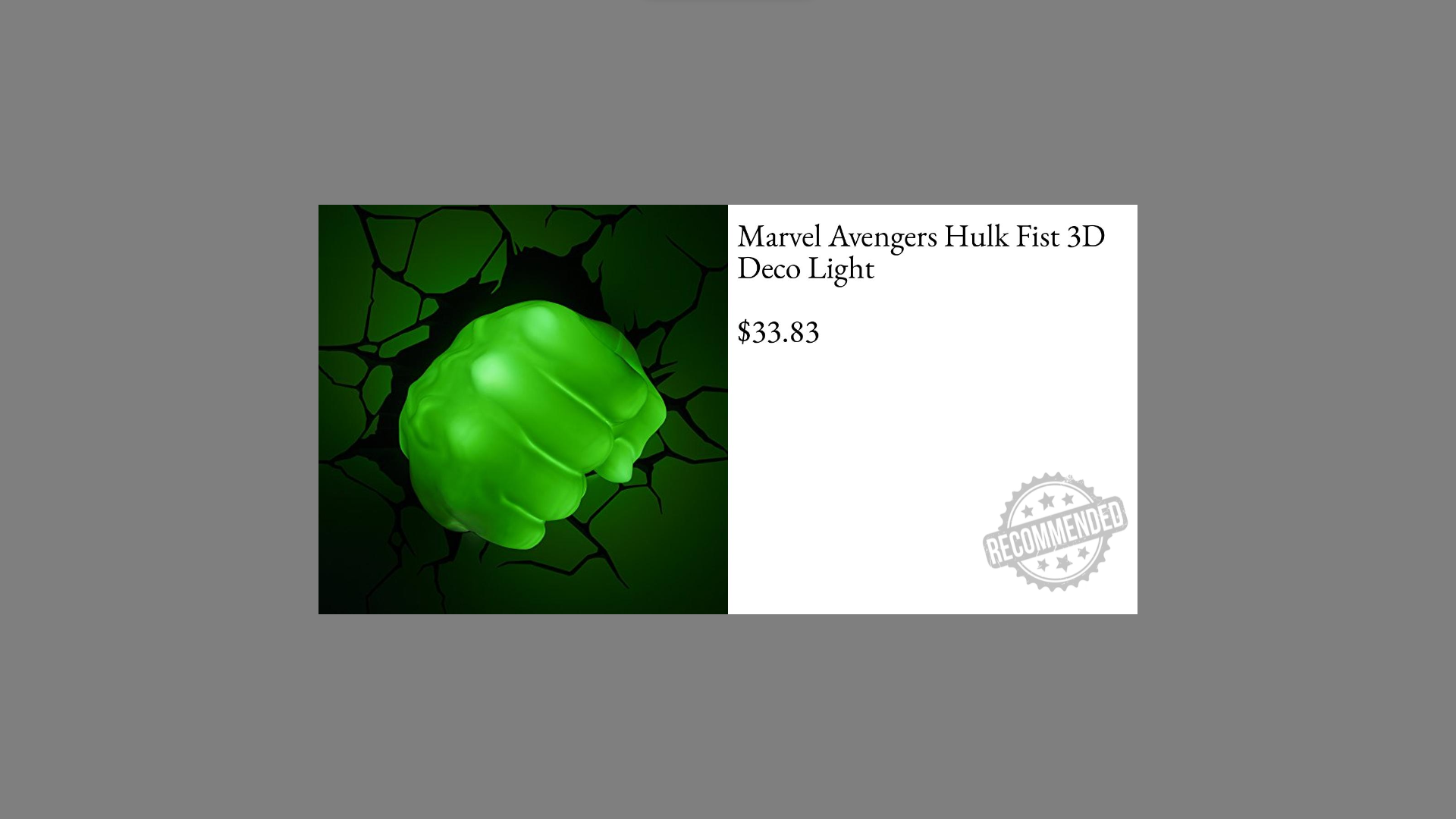}}
    }
    \hspace{0.0025\linewidth}
    \subfloat[Substitute Product]{
        \fbox{\includegraphics[trim={320 205 320 205}, clip, width=\w]{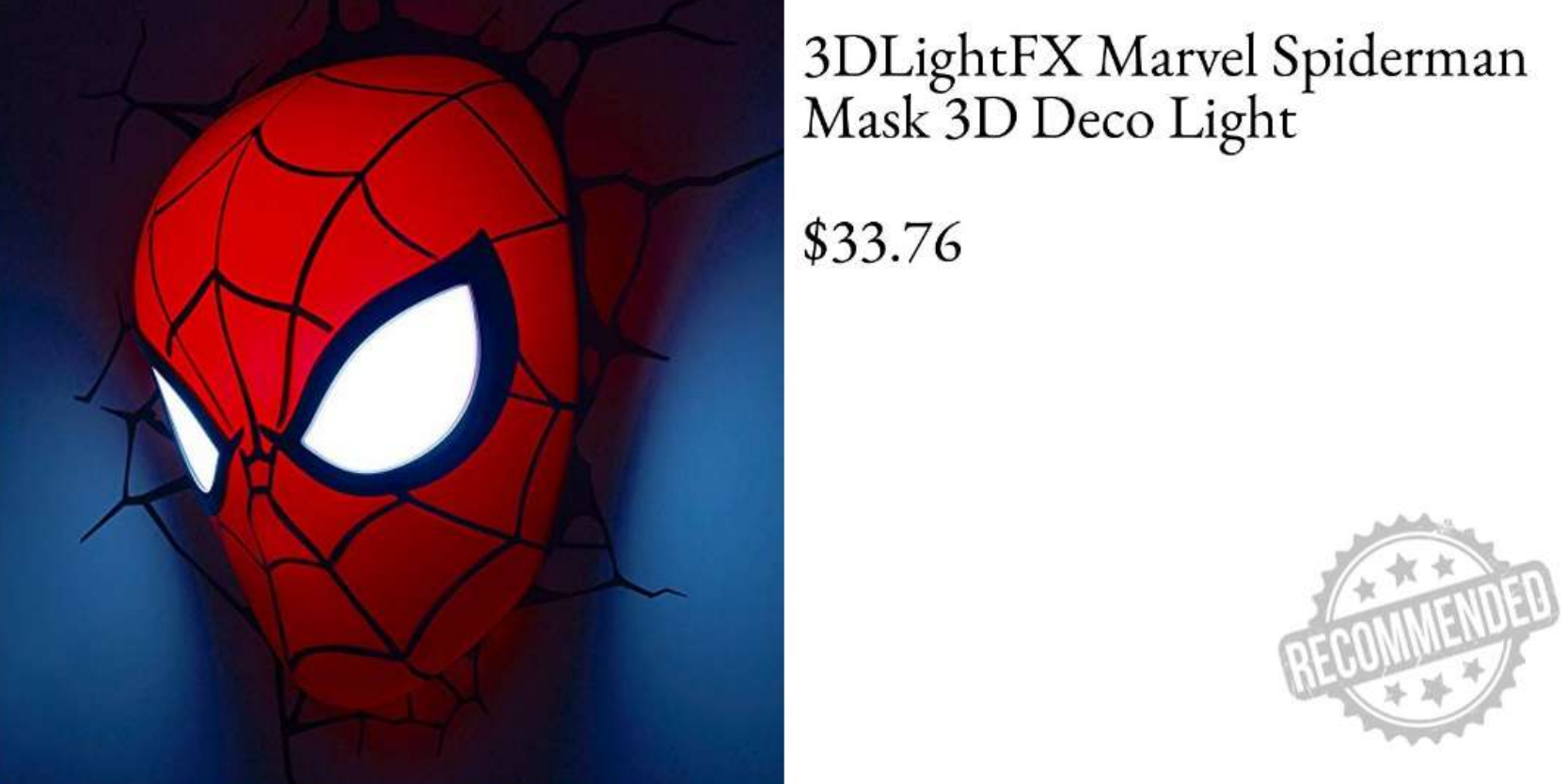}}
    }
    \hspace{0.0025\linewidth}
    \subfloat[Complement Product]{
        \fbox{\includegraphics[trim={320 205 320 205}, clip, width=\w]{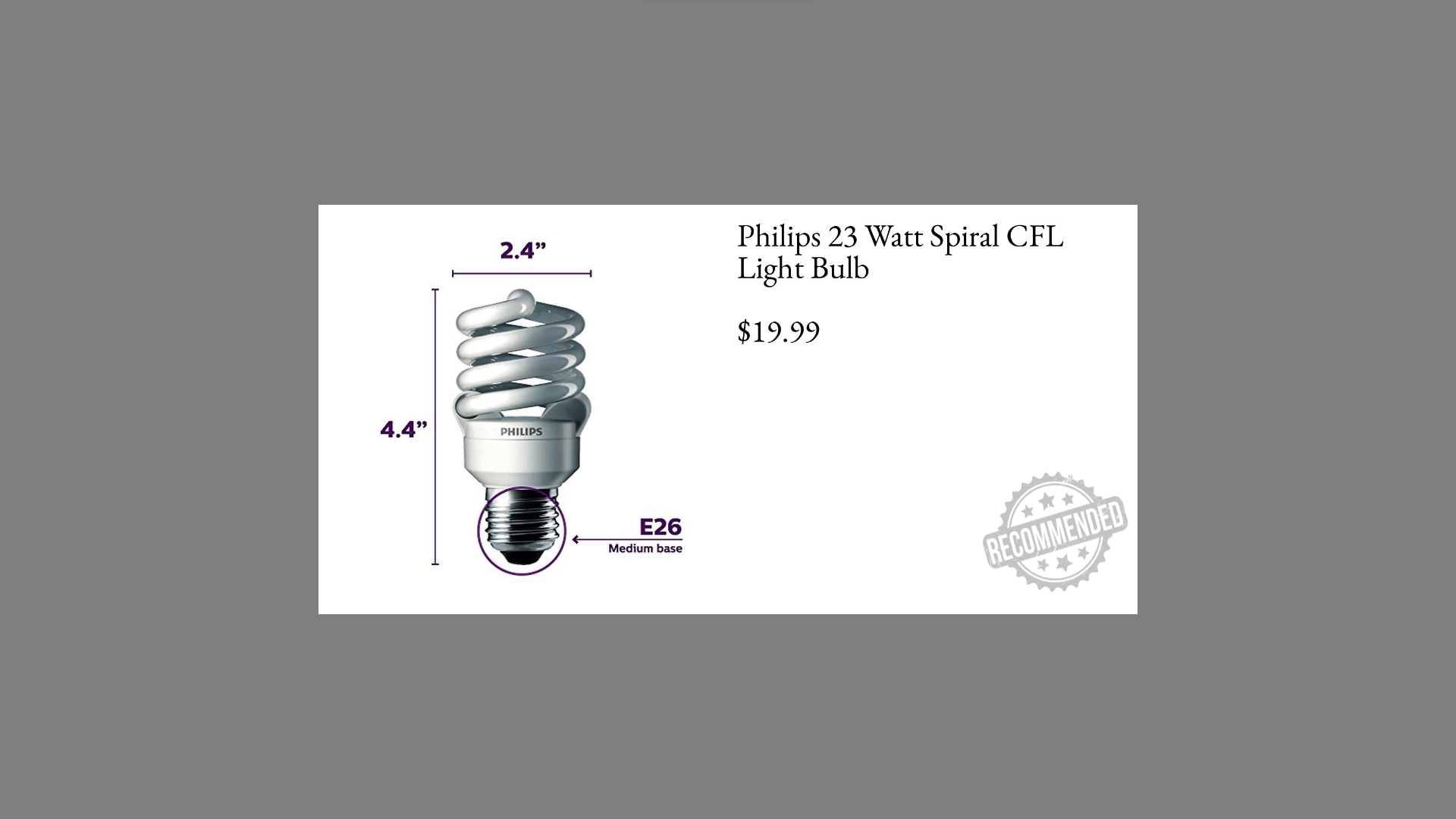}}
    }
    \hspace{0.0025\linewidth}
    \subfloat[Irrelevant Product]{
        \fbox{\includegraphics[trim={320 205 320 205}, clip, width=\w]{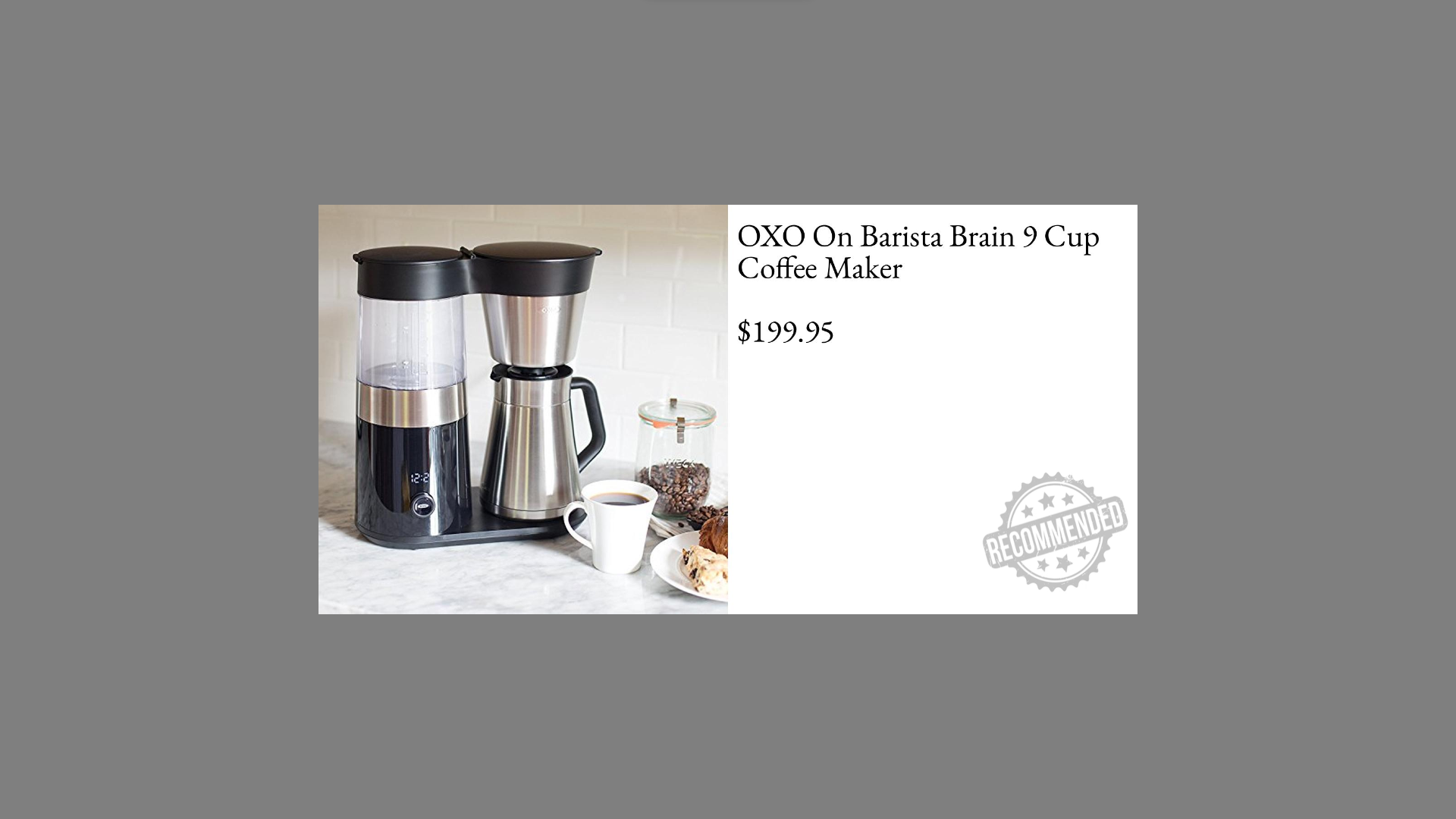}}
    }
    \caption{Stimuli examples: (a) Query ``Avengers night light''; (b) Exact recommendation; (c) Substitute recommendation (Marvel night light); (d) Complementary recommendation (light bulb that can be used with a night light); (e) Irrelevant recommendation.}
    \label{fig:stimuli}
\end{figure*}

\begin{table}[h]
\renewcommand{\arraystretch}{1.0}
\caption{Dataset statistics on Joint ESCI/Amazon Review Datasets, including item counts and percentage.}
\centering
\resizebox{0.9\linewidth}{!}{
\begin{tabular}{|l|r|r|}
\hline
\textbf{Dataset Content} & \textbf{Joint Dataset} & \textbf{Used in Study} \\
\hline
Dataset length in rows & 63,261 & 974 \\
Unique queries & 13,214 & 120 \\
Unique items & 41,781 & 353 \\
\hline
\multicolumn{3}{|c|}{\textbf{ESCI Label Distribution}} \\
\hline
Exact & 52,104 (82.36\%) & 263 (74.50\%)\\
Substitute & 7,519 (11.89\%) & 30 (8.5\%)\\
Irrelevant & 2,507 (3.96\%) & 30 (8.5\%)\\
Complement & 1,131 (1.79\%) & 30 (8.5\%)\\
\hline
\multicolumn{3}{|c|}{\textbf{Top 12 Main Categories by Unique Items}} \\
\hline
Amazon Home & 10,005 (23.95\%) & 69 (19.55\%) \\
Tools \& Home Improvement & 4,811 (11.51\%) & 33 (9.35\%) \\
Toys \& Games & 4,437 (10.62\%) & 57 (16.15\%) \\
Sports \& Outdoors & 4,138 (9.90\%) & 61 (17.28\%) \\
Grocery & 3,844 (9.20\%) & 27 (7.65\%) \\
Automotive & 3,807 (9.11\%) & 7 (1.98\%) \\
Amazon Fashion & 2,789 (6.68\%) & 23 (6.52\%) \\
Office Products & 2,571 (6.15\%) & 42 (11.90\%) \\
Pet Supplies & 2,292 (5.49\%) & 18 (5.10\%) \\
Industrial \& Scientific & 1,075 (2.57\%) & 10 (2.83\%) \\
Arts, Crafts \& Sewing & 1,059 (2.53\%) & 2 (0.57\%) \\
Computers & 953 (2.28\%) & 4 (1.13\%) \\
\hline
\end{tabular}}
\label{tab:esci_datasets}
\end{table}

\subsubsection{Ground-truth Labels}
\label{sssec:ground_truth_labels}

To determine the recommendation category of each product with respect to a query, we applied the ESCI relevance judgements, as specified by \citet{mcauley2015}:

\begin{itemize}
    \item \ul{\textbf{E}xact:} the item is relevant for the query, and satisfies all the query specifications (e.g., a water bottle matching all attributes of a query ``plastic water bottle 24oz'', such as material and size).
    \item \ul{\textbf{S}ubstitute:} the item is somewhat relevant, i.e., it fails to fulfil some aspects of the query but the item can be used as a functional substitute (e.g., fleece for a ``sweater'' query)
    \item \ul{\textbf{C}omplement:} the item does not fulfill the query, but could be used in combination with an exact item (e.g., track pants for ``running shoes'' query)
    \item \ul{\textbf{I}rrelevant:} the item is irrelevant, or it fails to fulfill a central aspect of the query (e.g., socks for a ``telescope'' query, or a wheat flour bread for a ``gluten–free bread'' query)
\end{itemize}

\subsubsection{Stimuli}
\label{sssec:stimuli}

From the joint dataset (\Cref{tab:esci_datasets}: ``Used in Study''), we sampled 120 queries in a uniform way across product categories (e.g., books, sports, electronics, clothing, etc.) and paired them with 120 product recommendations, 30 per recommendation category (e.g., \textbf{E}xact, \textbf{S}ubstitute). \textcolor{black}{The use of search queries allows us to capture user preferences and simulate interaction history—signals commonly used in \gls{rs}. This information can subsequently inform the ranking of items in a recommendation setting, thereby aligning our experimental findings with real-world applications.} 

Between the query and the product recommendation, we introduced a sequence of $k\in\{1,2,3\}$ \textbf{E}xact products, thus varying the position of the recommended product between second, third, or fourth in the sequence. This afforded a more ecologically valid setting---similar to the oddball paradigm \cite{sutton1965evoked}---where the participant is not conditioned to expect the recommendation at a fixed position. Each product description was shortened to 10–15 words using Meta's Llama 3 model via the LM Studio API. A custom prompt was used to \textit{consistently} instruct the model to retain only the most useful and descriptive information relevant to the query, removing unnecessary details (e.g., product codes). We further used the EB Garamond font to optimise for reading speed \cite{Wallace2022} and reserve more time for cognitive processing of the stimuli. When more than one product image was available, we opted for the most representative using a majority voting based on three human annotators. For the stimuli design we applied a template that resembled a simplified version of Amazon e-commerce platform, showing the product image on the left side of the visual box and the product description and price to the top right (\Cref{fig:stimuli}). A subtle ``Recommended'' logo was added for recommended products, to suggest their condition.

\subsection{Procedure}
\label{ssec:procedure}

Prior to the experiment, participants were asked to complete an entry questionnaire that collected information related to exclusion criteria (e.g., vision issues or neurological difficulties) and prior experience with e-commerce platforms. Then, a short tutorial and Q/A session was held that covered the study in detail, during which participants were informed about their right to withdraw from the experiment at any moment without facing any adverse consequences. Upon signing a consent form, a preparation of the equipment was made (i.e. fitting the \gls{eeg} cap, applying the gel to the electrodes, checking the impedance level for each channel). 

\begin{figure*}[t!]
    \begin{center}
    \includegraphics[width=0.86\textwidth]{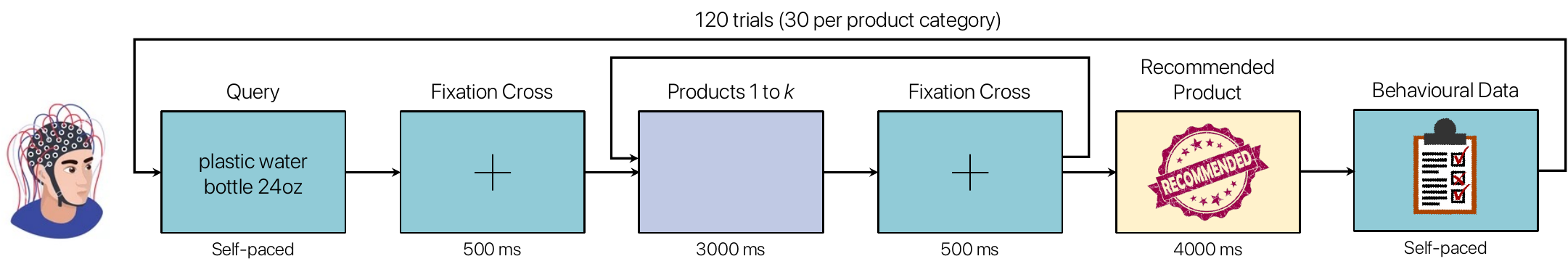}
    \end{center}
    \caption{Overview of single‐trial experimental protocol. Each participant performs 120 trials. In each trial, the participant is shown a query, followed by a sequence of $k\in\{1,2,3\}$ exact products, and concluded by a product recommendation that can be any of the four categories (Exact, Substitute, Complement, or Irrelevant).}
    \label{fig:protocol}
\end{figure*}

Our experiment consisted of 120 trials that implemented the oddball paradigm \cite{sutton1965evoked}. In each trial, participants were presented with a search query followed by a sequence of one to three \textbf{E}xact products for context, followed by a product recommendation that belonged to one of four recommendation categories (\Cref{fig:protocol}). The presentation duration was set to 3 seconds for context products and 4 seconds for the recommendation. At the end of each trial, participants had to self-label the recommended product and rate it according to the following dimensions: (1) Relevance, (2) Likelihood of purchase, and (3) Diversity. Ratings were assessed using a 5-point Likert scale (higher values indicate a stronger agreement). 

To control for order effects, all participants viewed the same set of samples in a randomized order, following a Latin-squares design. Also, both query reading and product ratings were self-paced (to ensure task comprehension), and there was no time limit for completing each trial. Considering \gls{eeg} equipment constraints and potential participant fatigue, we limited the experiment duration to 90-120 min and incorporated breaks at the one-third and two-thirds marks. Upon completing the study, participants were asked to complete an exit questionnaire that inquired about demographics. 

\begin{figure}[t!]
    \begin{center}
    \includegraphics[width=0.75\linewidth]{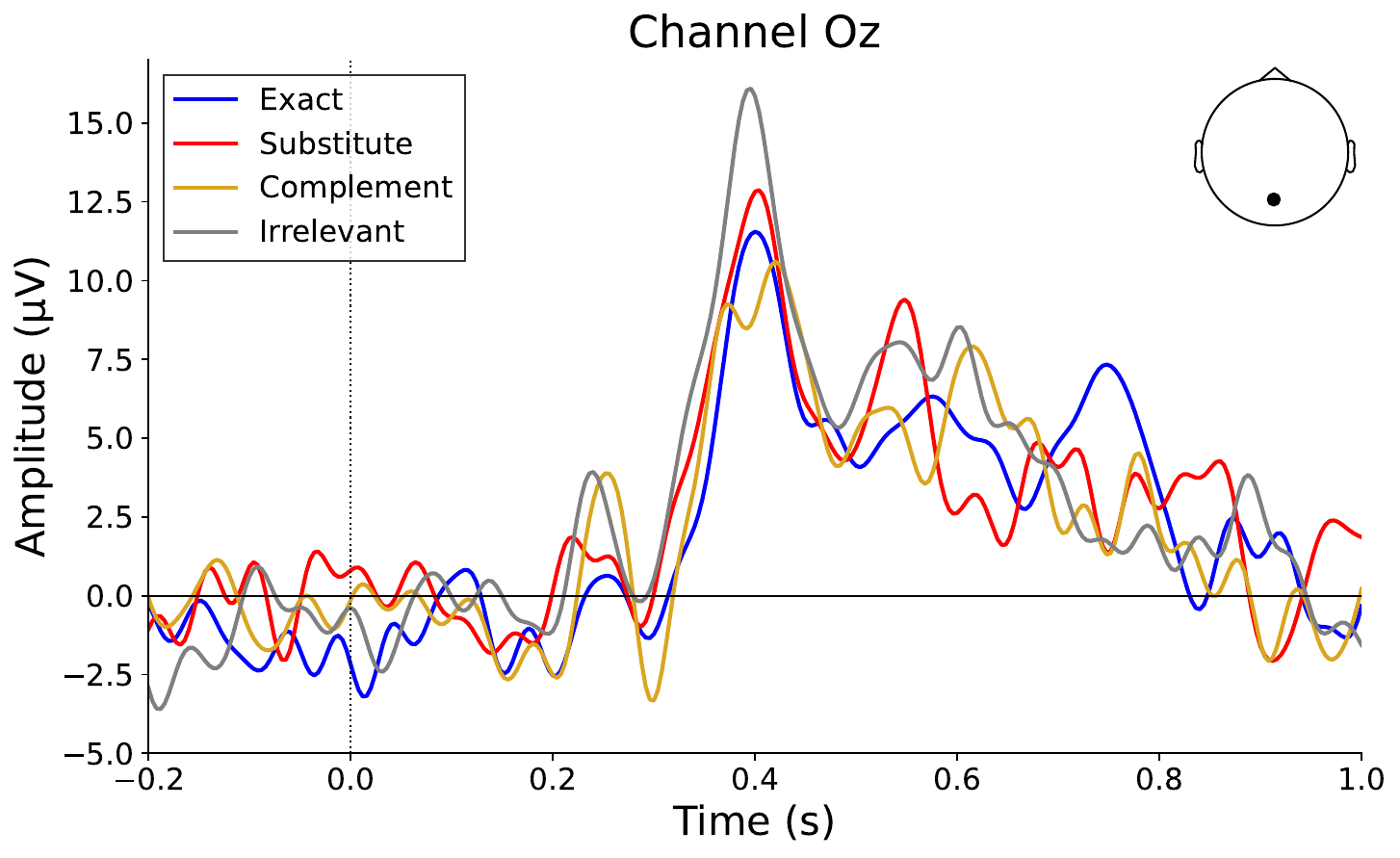}
    \end{center}
    \caption{Example \gls{erp} showing average neural responses to four recommendation categories}
    \label{fig:erp}
\end{figure}

\subsection{Behavioural analysis}
\label{ssec:behavioural_analysis}

To assess participants' decision-making and product categorization accuracy (i.e., how well self-labels matched the original dataset labels) we performed a thorough analysis of behavioural data, focusing on classification accuracy, relevance ratings, purchase likelihood, and diversity. Additionally, we investigated the statistical relationships between these measures to understand behavioural patterns in the task. Specifically, we conducted multiple statistical tests: the Shapiro-Wilk test was used to assess normality, Levene’s test examined homogeneity of variance, while classification accuracy was determined using Chi-square tests. Depending on the aforementioned tests
, the Friedman test was applied for comparing multiple related conditions, and the Wilcoxon signed-rank test was used for pairwise comparisons of behavioural measures  

\subsection{\gls{eeg} pre-processing}
\label{ssec:eeg_preprocessing}

We excluded two participants (out of 21) from subsequent analyses due to poor task comprehension and low \gls{snr}. \gls{eeg} data were processed using the MNE-Python library \cite{GramfortEtAl2013a}.
We extracted 4.5-second epochs around each event, including a 0.5-second pre-stimulus baseline for correction, per trial. Data were band-pass filtered between 0.5-20 Hz, as our initial spectral analysis showed no stimulus-specific information above 20 Hz. Artifact rejection was performed using FASTER \cite{nolan2010faster}, an automated de-noising pipeline that 
handles channel rejection, epoch removal, \gls{ica}, and channel interpolation. The outcome of this pre-processing is the Single-trial \glspl{erp} computed per epoch (
\Cref{fig:erp}).

\subsection{\gls{eeg} analysis}
\label{ssec:eeg_analysis}

In what follows, we present a comprehensive analysis of the \gls{eeg} data through three complementary approaches. First, we conduct within-subjects classification to examine individual-specific neural patterns and their variability across recommendation categories. Second, we perform between-subjects analysis to identify common neural signatures and evaluate the generalisability of our findings across participants. Finally, we analyse established neurophysiological markers of engagement to quantify how different recommendation categories modulate user responses.

\subsubsection{Within-subjects classification}
\label{sssec:within_subjects_classification}

We investigated whether distinct neural patterns emerge across recommendation categories at the participant level. In line with prior work \cite{Tukka,lotte2018review,wei2016classification,subasi2017classification,zhang2020eeg,srivastava2013support}, we trained a \gls{svm} classifier to perform pairwise classification between recommendation categories, using \gls{eeg} features supported by prior research \cite{eegKl,buzsaki2006rhythms,buzsaki2012high,Tang2024}. Specifically, the classifier's input consisted of several features derived from each epoch: (1) Single-trial \glspl{erp}, representing the denoised temporal response; (2) \gls{psd} within relevant frequency bands, using the Welch's method; (3) \gls{kl} relative to the mean training-evoked responses \cite{eegKl,zhang2024eeggan}; (4) signal complexity measures. All features were computed over a one-second window following stimulus presentation. To reduce input dimensionality, \gls{pca} was applied, retaining 99\% of the variance. \textcolor{black}{We omit multi-class classification due to a well-known issue of performance degradation \cite{Yang2025,Abdulkarim2021,LindigLeon2020}.}

\glspl{erp} capture the early neural response to a stimulus, identifiable by a distinct voltage peak. Their latency, topography, and amplitude correlate with specific cognitive functions, making \glspl{erp} a useful metric for studying neural processes.  Complementing this, our frequency-based features target well-established \gls{eeg} bands: delta (0.5–4 Hz), theta (4–8 Hz), alpha (8–12 Hz), and low-beta (13–20 Hz) \cite{buzsaki2006rhythms}. Each band is associated with distinct cognitive processes: delta with attention and cognitive effort, theta with memory encoding and cognitive control, alpha with attentional suppression and workload regulation, and low-beta with active thinking and decision-making. Our preliminary spectral analysis indicated that frequencies above 20 Hz exhibited minimal task-related variation, prompting the exclusion of high-beta and gamma bands to reduce muscle artifact influence \cite{buzsaki2012high}.

Additional features were derived through data-driven methods. For example, \gls{kl} divergence was used to quantify the similarity between signals within the same condition, while complexity measures like approximate entropy \cite{Zuniga2024,Pappalettera2023}, Higuchi's fractal dimension \cite{Zuniga2024,Tang2024,IBANEZMOLINA201469}, Katz's fractal dimension \cite{Tang2024,904882}, and \gls{dfa} \cite{Tang2024,PhysRevE.49.1685}, helped differentiate experimental conditions based on their inherent signal complexity \cite{Tang2024}.

To ensure stable performance estimates, we created five different train-test splits (70\% training, 30\% testing) from the initial dataset. For each split, the classes were balanced via random down-sampling. The down-sampling was done five times per split to avoid biases, yielding 25 dataset variations per participant for independent performance evaluation and accounting for data variability. 

\begin{table*}[t!]
\caption{\textcolor{black}{Classification accuracy (Acc.) and \% of difference (Diff) from 
baseline; *$p{<}.05$, **$p{<}.01$, ***$p{<}.001$ (Bonferroni-corrected).}}
\centering
\setlength{\tabcolsep}{4pt}
\resizebox{0.99\linewidth}{!}{
\begin{tabular}{|l|cc|cc|cc|cc|cc|cc|}
\hline
\multirow{2}{*}{\textbf{Feature set}} 
& \multicolumn{2}{c|}{\textbf{Exact vs. Complement}} 
& \multicolumn{2}{c|}{\textbf{Exact vs. Substitute}} 
& \multicolumn{2}{c|}{\textbf{Exact vs. Irrelevant}} 
& \multicolumn{2}{c|}{\textbf{Complement vs. Substitute}} 
& \multicolumn{2}{c|}{\textbf{Complement vs. Irrelevant}} 
& \multicolumn{2}{c|}{\textbf{Substitute vs. Irrelevant}} \\
\cline{2-13}
& Acc. & Diff (\%) & Acc. & Diff (\%) & Acc. & Diff (\%) & Acc. & Diff (\%) & Acc. & Diff (\%) & Acc. & Diff (\%) \\
\hline
EEG Theta Delta      & 0.545*** & 9.0  & 0.524*** & 4.8  & 0.545*** & 9.0  & 0.519*  & 3.8  & 0.531*** & 6.2  & 0.537*** & 7.4  \\
EEG KL               & 0.545*** & 9.0  & 0.524*** & 4.8  & 0.544*** & 8.8  & 0.519*  & 3.8  & 0.530*** & 6.0  & 0.537*** & 7.4  \\
EEG DFA HFD          & \textbf{0.546}*** & \textbf{9.2}  & 0.524*** & 4.8  & 0.544*** & 8.8  & 0.520*  & 4.0  & 0.529*** & 5.8  & 0.537*** & 7.4  \\
EEG KL Delta         & 0.544*** & 8.8  & \textbf{0.525}*** & \textbf{5.0}  & 0.544*** & 8.8  & 0.518*  & 3.6  & 0.530*** & 6.0  & 0.537*** & 7.4  \\
EEG Theta HFD        & \textbf{0.546}*** & \textbf{9.2}  & 0.524*** & 4.8  & \textbf{0.546}*** & \textbf{9.2}  & 0.519*  & 3.8  & 0.530*** & 6.0  & 0.537*** & 7.4  \\
KL Beta              & 0.496    & -0.8 & 0.495    & -1.0 & 0.511    & 2.2  & \textbf{0.533}*** & \textbf{6.6}  & 0.509    & 1.8  & 0.509    & 1.8  \\
Entropy Theta Delta  & 0.505    & 1.0  & 0.483    & -3.4 & 0.539*** & 7.8  & 0.483    & -3.4 & \textbf{0.549}*** & \textbf{9.8}  & 0.543*** & 8.6  \\
KL Theta Delta       & 0.503    & 0.6  & 0.492    & -1.6 & 0.539*** & 7.8  & 0.495    & -1.0 & 0.535*** & 7.0  & \textbf{0.549}*** & \textbf{9.8}  \\
\hline
\end{tabular}
}
\label{tab:eeg_comparison_accuracy_1}
\end{table*}

\begin{figure*}[t!]
    \begin{center}
    \includegraphics[width=0.7\linewidth]{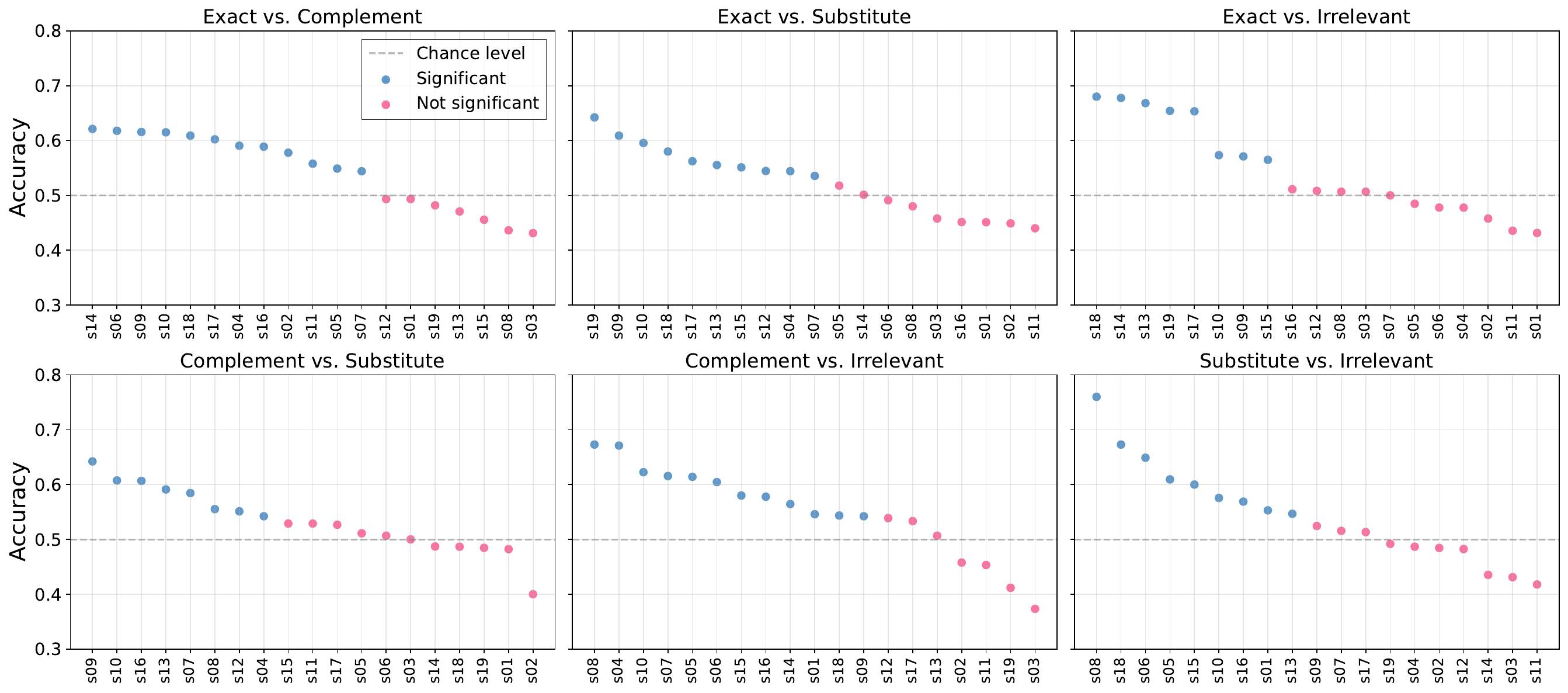}
    \end{center}
    \caption{Accuracy plots of the best performing feature combination per condition, according to \Cref{tab:eeg_comparison_accuracy_3}.}
    \label{fig:acc_plots}
\end{figure*}

\subsubsection{Between-subject classification}
\label{sssec:between_subject_classification}

We expanded our experiments to explore and characterise consistent neural patterns across participants. Given the inherent noise in raw \gls{eeg} signals, group-level analysis is particularly advantageous as it enables the use of \glspl{erp} for such a purpose.

The typical approach to \gls{erp} computation involves averaging all epochs per condition and participant, which reduces our dataset from (N = 19 participants $\times$ 30 epochs $\times$ 4 conditions) to (N = 19 subjects $\times$ 1 \glspl{erp} $\times$ 4 conditions). Moreover, given that each participant contributed a moderate number of epochs per condition, this can lead to suboptimal \glspl{erp}. To address these limitations, we implemented a data augmentation strategy that generates multiple synthetic \glspl {erp} (N = 19 participants $\times$ 50 \glspl{erp} $\times$ 4 categories) per participant and condition through a pseudo-random linear combination of epochs. 
Specifically, the variability of each epoch was used to weight its contribution to the synthetic \glspl{erp}, with larger variability assumed to indicate broader coverage of brain activity. To further increase diversity, these weights were adjusted by introducing noise, multiplying them by a random factor within the range [0.5, 1]. This approach was designed to model uncontrolled variability in attention and brain activity during each task. The adjusted weights were then used in a weighted averaging process to generate synthetic \glspl{erp}, which were subsequently used for training.

Considering the potential noise in the derived \glspl{erp}, we implemented a data-driven classification approach for each pair of conditions, using a two-stage cross-validation approach. First, an outer leave-one-out cross-validation was employed, splitting the participants and all their \gls{erp} variations into training and test sets. Within each training set, an inner 5-fold cross-validation was used to train a base model for each \gls{eeg} channel. Each model consisted of a \gls{mlp} with a single hidden layer of 50 units. Training was limited to 500 iterations, with a regularisation parameter $\alpha = 0.05$ to prevent overfitting.

The mean validation accuracy of each base model was then used to generate a topomap (\textcolor{black}{\autoref{fig:BaseModels}}), illustrating the predictive power of each electrode's recorded activity. Additionally, the probabilistic predictions from these models served as features for training and testing a logistic regression classifier, at the outer cross-validation level. This meta-model predicted the final accuracy for each subject based on its predictions. Finally, the mean accuracy across all participants was computed. Given the stochastic nature of \gls{erp} generation, the entire process was repeated five times, and the mean and standard deviation of the resulting accuracies were computed.

\subsubsection{Between-subjects engagement analysis}
\label{sssec:between_subjects_engagement_analysis}

\gls{eeg} data were analysed to examine \gls{faa}, a proxy of engagement, approach-avoidance motivation \cite{Thibodeau2006}, and emotional regulation \cite{harmon2010role}. Specifically, \gls{faa} is calculated as the difference in alpha power between left (F3) and right (F4) frontal electrodes, which has been associated with emotional processing, cognitive control, and decision-making \cite{gotlib1998eeg}: 
\begin{equation}
    \label{eq:1}
    \text{FAA} = \log(P_{F4}) - \log(P_{F3})
\end{equation}

Higher \gls{faa} values indicate greater relative right-frontal activity, which has been linked to withdrawal-related emotions, while lower (or negative) values indicate greater left-frontal dominance, often associated with approach motivation \cite{barros2022frontal}. Additionally, theta and beta frequency bands' power was analysed following prior work \cite{halderman2021eeg}. In a similar way, higher powers indicate a higher engagement of the participant with the task, providing complementary insight into neural processing during the experiment.

To extract the different frequency bands' power, we applied Welch's method to compute \gls{psd} within the specific ranges as described previously. Welch's method is commonly used in \gls{eeg} research to estimate power in different frequency bands, as it improves the robustness of spectral estimates by averaging over multiple overlapping windows \cite{barbe2009welch}. The resulting \gls{psd} values for the electrodes of interest were averaged across the epoch to obtain a single mean power value per electrode (C3 for theta/beta bands and F3/F4 for \gls{faa} calculation). For this analysis, six subjects were excluded due to poor \gls{snr} for the corresponding calculations.

\begin{table*}[t!]
\caption{\textcolor{black}{Mean classification accuracy (Acc.) and \% of difference (Diff) from random baseline (only above-chance results from \Cref{tab:eeg_comparison_accuracy_1} as reported by \citet{Tukka}) for each feature set and recommendation category comparison.
}}
\centering
\setlength{\tabcolsep}{4pt}
\resizebox{0.99\linewidth}{!}{
\begin{tabular}{|l|cc|cc|cc|cc|cc|cc|}
\hline
\multirow{2}{*}{\textbf{Feature set}} 
& \multicolumn{2}{c|}{\textbf{Exact vs. Complement}} 
& \multicolumn{2}{c|}{\textbf{Exact vs. Substitute}} 
& \multicolumn{2}{c|}{\textbf{Exact vs. Irrelevant}} 
& \multicolumn{2}{c|}{\textbf{Complement vs. Substitute}} 
& \multicolumn{2}{c|}{\textbf{Complement vs. Irrelevant}} 
& \multicolumn{2}{c|}{\textbf{Substitute vs. Irrelevant}} \\
\cline{2-13}
& Acc. & Diff (\%) & Acc. & Diff (\%) & Acc. & Diff (\%) & Acc. & Diff (\%) & Acc. & Diff (\%) & Acc. & Diff (\%) \\
\hline
EEG Theta Delta      & \textbf{0.591} & \textbf{18.2} & \textbf{0.567} & \textbf{13.4} & 0.583 & 16.6 & \textbf{0.570} & \textbf{14.0} & 0.580 & 16.0 & 0.562 & 12.4 \\
EEG KL               & \textbf{0.591} & \textbf{18.2} & \textbf{0.567} & \textbf{13.4} & 0.590 & 18.0 & 0.563 & 12.6 & \textbf{0.581} & \textbf{16.2} & 0.558 & 11.6 \\
EEG DFA HFD          & \textbf{0.591} & \textbf{18.2} & 0.566 & 13.2 & 0.590 & 18.0 & 0.569 & 13.8 & 0.579 & 15.8 & 0.558 & 11.6 \\
EEG KL Delta         & \textbf{0.591} & \textbf{18.2} & \textbf{0.567} & \textbf{13.4} & 0.597 & 19.4 & 0.568 & 13.6 & 0.580 & 16.0 & 0.558 & 11.6 \\
EEG Theta HFD        & 0.590 & 18.0 & 0.565 & 13.0 & \textbf{0.599} & \textbf{19.8} & \textbf{0.570} & \textbf{14.0} & 0.579 & 15.8 & 0.559 & 11.8 \\
KL Beta              & 0.562 & 12.4 & 0.549 & 9.8 & 0.556 & 11.2 & 0.560 & 12.0 & 0.558 & 11.6 & 0.547 & 9.4 \\
Entropy Theta Delta  & 0.568 & 13.6 & 0.546 & 9.2 & 0.577 & 15.4 & 0.559 & 11.8 & \textbf{0.582} & \textbf{16.4} & \textbf{0.591} & \textbf{18.2} \\
KL Theta Delta       & 0.572 & 14.4 & 0.540 & 8.0 & 0.577 & 15.4 & 0.547 & 9.4 & 0.576 & 15.2 & 0.579 & 15.8 \\
\hline
\end{tabular}
}
\label{tab:eeg_comparison_accuracy_3}
\end{table*}

\section{Results and Discussion}
\label{sec:results_and_discussion}

\subsection{Neural variation across recommendation categories}
\label{ssec:}

This section addresses \textbf{RQ1:} \textit{Do different recommendation categories evoke distinct neural signatures in brain activity as measured through \gls{eeg}, and can these patterns be reliably identified?} Initially, we report results on within-subjects classification that examines individual-specific neural patterns and their variability across recommendation categories. Then, we discuss our between-subjects analysis with aims to identify common neural signatures and evaluate the generalisability of our findings across participants. 

\subsubsection{Within-subject experiments}
We begin our analysis by focusing on the within-subject classifier performance to examine the consistency of neural responses to each recommendation category at the individual level. \Cref{tab:eeg_comparison_accuracy_1} shows the best performing configurations in terms of mean accuracy, for each pairwise comparison and across all participants. Our results suggest that, in most cases, the classifier's average performance exceeds chance levels, indicating the presence of meaningful neurophysiological differences between conditions \cite{Tukka}. These findings are further supported by a Mann-Whitney significance test, which confirms that our predictions significantly differ from chance (p < .01, Bonferroni corrected). 

Specifically, the ``E vs. S'' comparison is associated with the lowest mean accuracy, a trend that remains even after excluding individual models ($42\%$) with near-chance performance, as shown in \Cref{tab:eeg_comparison_accuracy_3} and \Cref{fig:acc_plots}. These results suggest that, while neural responses to \textbf{E}xact and \textbf{S}ubstitute items might still be distinguishable in some cases, the classifier's reduced performance in this pair suggests that these two categories consistently evoke more similar neural responses than other comparisons. This overlap highlights a key insight: from a neurological perspective, the brain may process substitute alternatives and exact matches in a closely related manner.

On the other hand, the ``C vs. I'' ($0.549$), ``S vs. I'' ($0.549$), and ``E vs. I'' ($0.546$) comparisons yielded the highest mean accuracies, demonstrating that the brain differentiates between product recommendations relevant to the search query (\textbf{E}xact, \textbf{C}omplement, or \textbf{S}ubstitute) and those that are not. This pattern holds after removing low-performing individual models, as shown in \Cref{tab:eeg_comparison_accuracy_1}. However, the ordering of accuracies shows minor variations across models, with ``E vs. I'' emerging as the top performer. This comparison also had the highest number of excluded subjects (\Cref{fig:acc_plots}), suggesting that although $42\%$ of individuals seem to display a clearer neural differentiation between exact and irrelevant items, a substantial subset ($58\%$) may not, reflecting a slightly larger inter-individual variability. Similar accuracies were observed for the ``E vs. C'' comparison (0.546 for the whole sample, and 0.591 after excluding $37\%$ of individual models, based on \Cref{fig:acc_plots}), indicating the presence of relevant brain activity differences. While speculative, these differences could be attributed to the associative and semantic processing involved in identifying complementary products, as opposed to the more straightforward processing of exact product matches.
 
\begin{table}[t!]
    \caption{Meta-classifier accuracy across different comparisons over multiple runs, for the between-subject classification strategy. Bold denotes best run.}
    \centering
    \resizebox{\columnwidth}{!}{  
    \begin{tabular}{| l | c | c | c | c | c | c |}
        \hline
        \textbf{Comparison} & \textbf{1} & \textbf{2} & \textbf{3} & \textbf{4} & \textbf{5} & \textbf{AVG $\pm$ STD} \\
        \hline
        E vs. I & \textbf{0.642} & 0.600 & 0.537 & 0.563 & 0.500 & 0.568 $\pm$ 0.055 \\
        E vs. C & \textbf{0.595} & 0.568 & 0.558 & 0.505 & 0.563 & 0.558 $\pm$ 0.033 \\
        E vs. S & 0.432 & 0.442 & 0.421 & 0.374 & \textbf{0.453} & 0.424 $\pm$ 0.031 \\
        I vs. C & 0.500 & \textbf{0.584} & 0.537 & 0.537 & 0.421 & 0.516 $\pm$ 0.061 \\
        I vs. S & 0.459 & 0.459 & 0.463 & 0.489 & \textbf{0.605} & 0.495 $\pm$ 0.063 \\
        C vs. S & 0.647 & \textbf{0.674} & 0.663 & 0.611 & 0.626 & 0.644 $\pm$ 0.026 \\
        \hline
    \end{tabular}
    }
    \label{tab:accuracy_metamodel}
\end{table}

Regarding the best-performing features in individual models, we observe that the raw \gls{eeg} signal consistently dominates over other features, leading to similar outcomes across all experiments that incorporate it. This suggests that the raw \gls{eeg} signal captures a substantial amount of information, making it a strong baseline for classification. However, for the top-performing comparisons, such as ``C vs. I'' and ``S vs. I'', combinations of a few derived parameters (most notably theta frequency power, which is also among the top-performing features in the ``E vs. I'' model) yielded better results. This suggests that these derived features are particularly well-suited to capturing distinctions between these specific recommendation categories. These findings align with those of \citet{Liang2017}, who demonstrated that frontal theta activity supports the detection of mismatched information in visual working memory tasks.

In other comparisons, models using derived features underperformed those based on raw \gls{eeg} signals, likely due to a loss of neural complexity and unmodelled data variations. Overall, within-subject results align with observed user behaviour (\Cref{ssec:rq2_results}). Above-chance classification across comparisons suggests distinct neural correlates, though expectedly overlapping categories (e.g., ``E vs. S'' and ``C vs. S'') show weaker differentiation.

\begin{figure*}[t!]
    \def\w{0.12\linewidth}
    \subfloat[\centering C vs. S]
    {
        \includegraphics[width=\w]{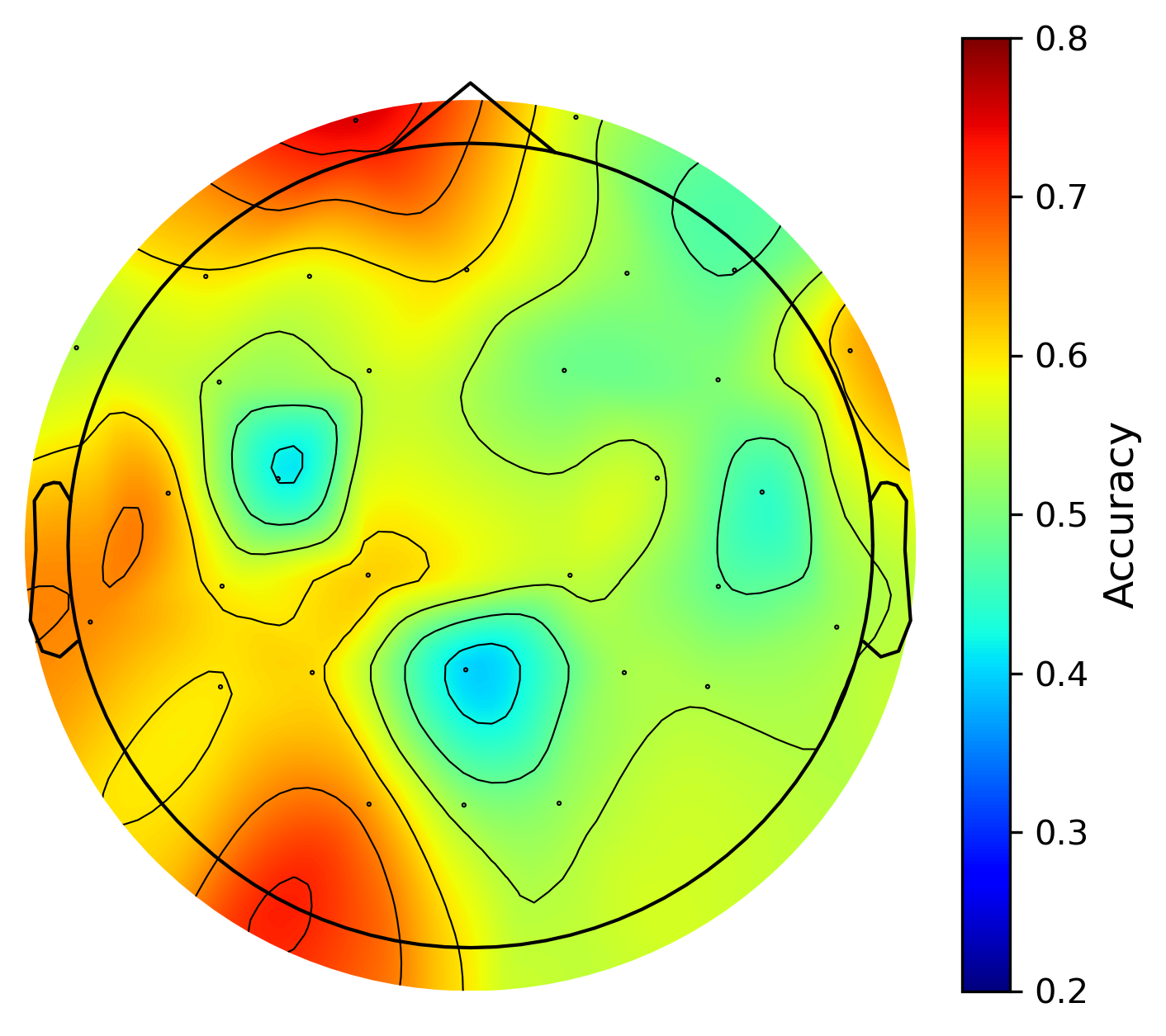}
    }
    \hspace{0.01\linewidth}
    \subfloat[\centering E vs. C]
    {
        \includegraphics[width=\w]{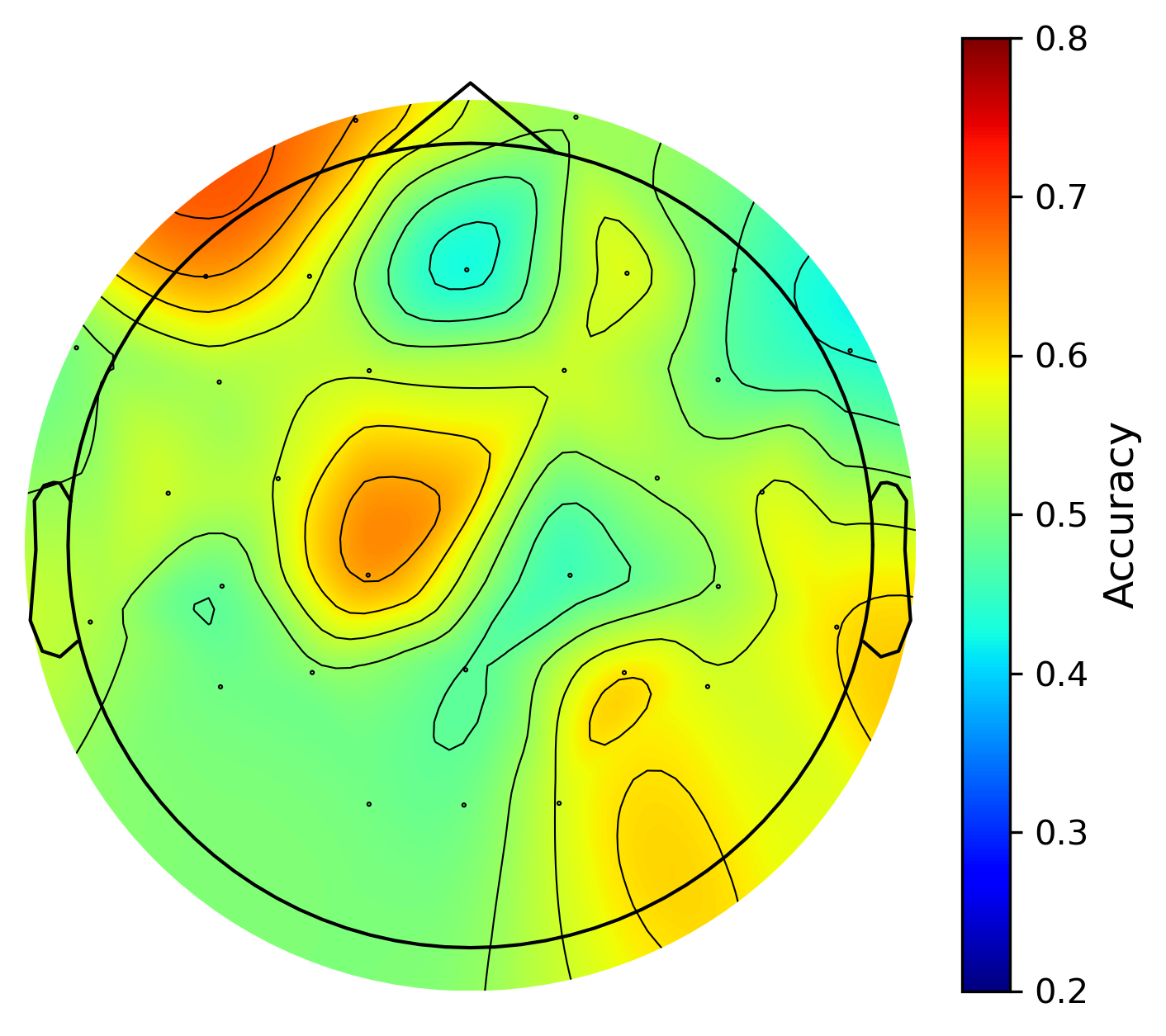}
    }
    \hspace{0.01\linewidth}
    \subfloat[\centering E vs. I]
    {
        \includegraphics[width=\w]{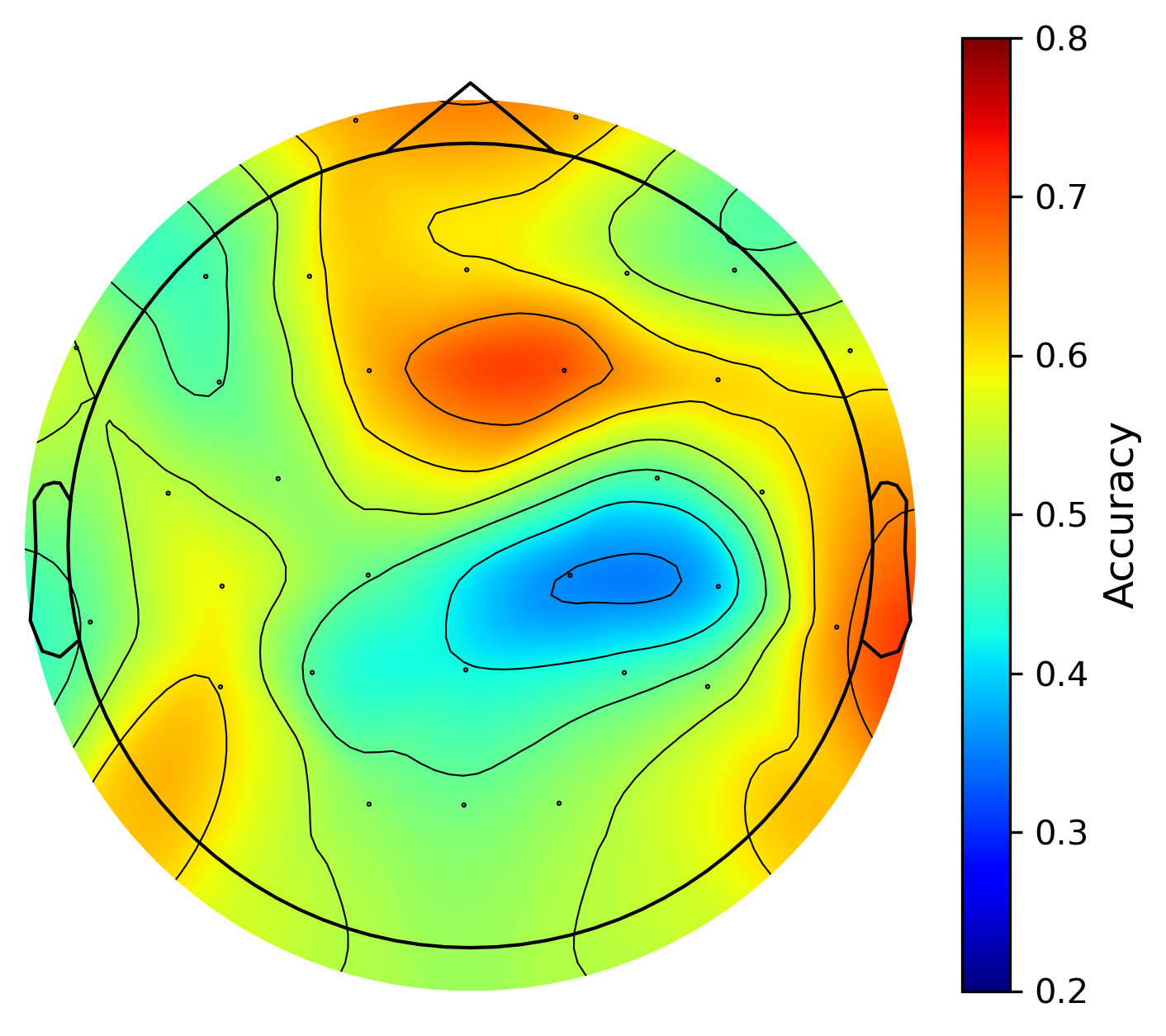}
    }
    \hspace{0.01\linewidth}
    \subfloat[\centering E vs. S]
    {
        \includegraphics[width=\w]{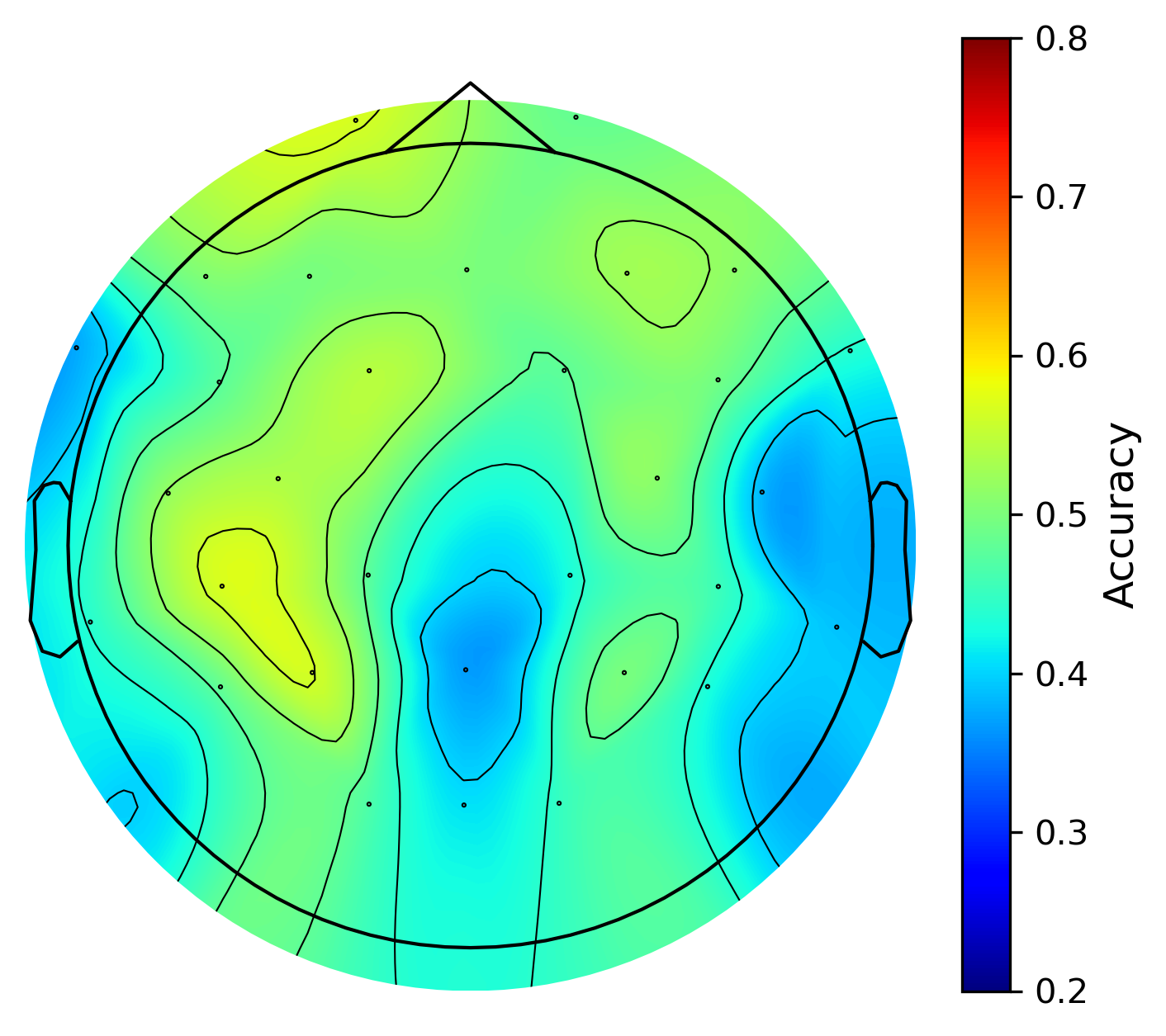}
    }
    \hspace{0.01\linewidth}
    \subfloat[\centering I vs. C]
    {
        \includegraphics[width=\w]{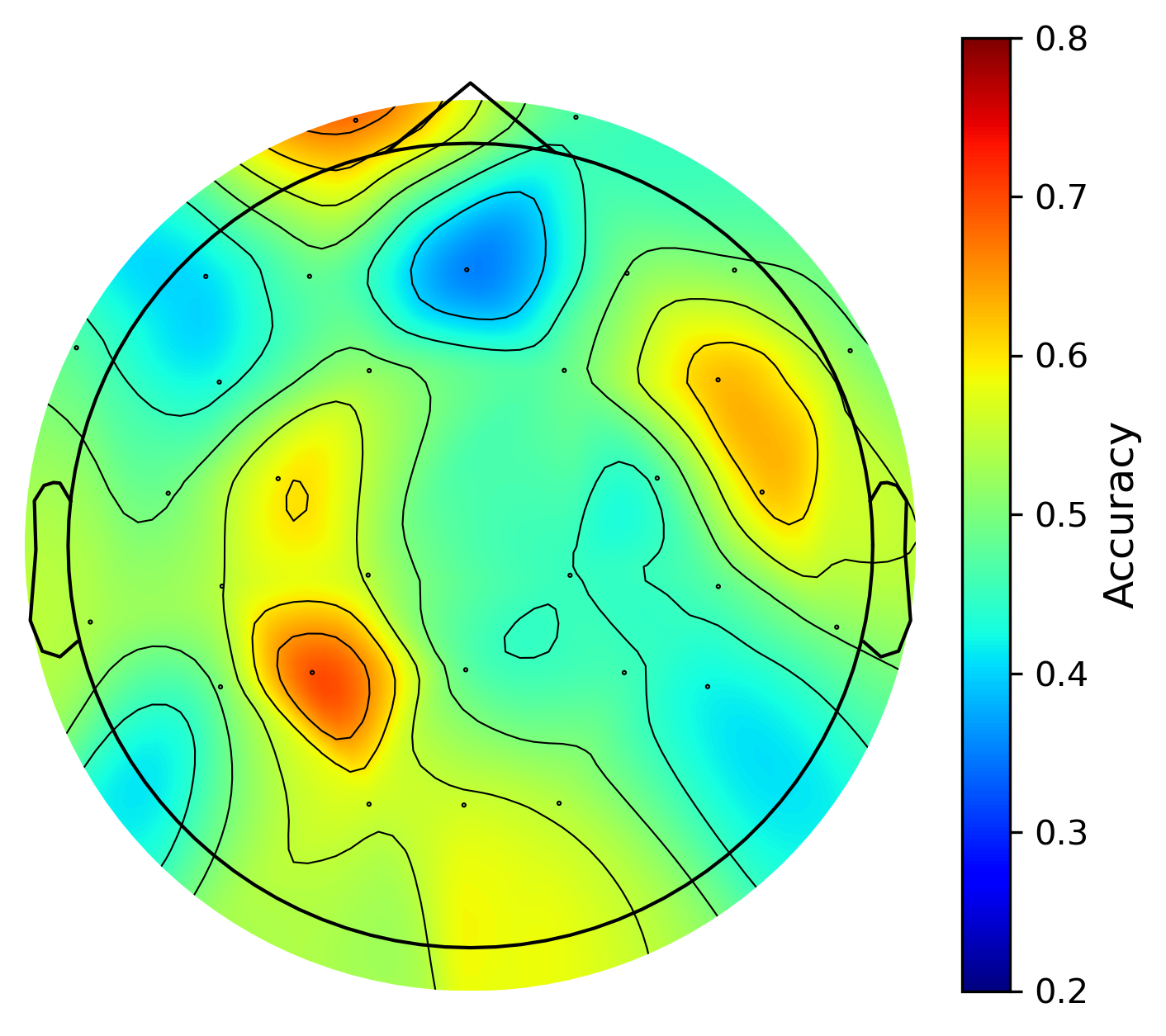}
    }
    \hspace{0.01\linewidth}
    \subfloat[\centering I vs. S]
    {
        \includegraphics[width=\w]{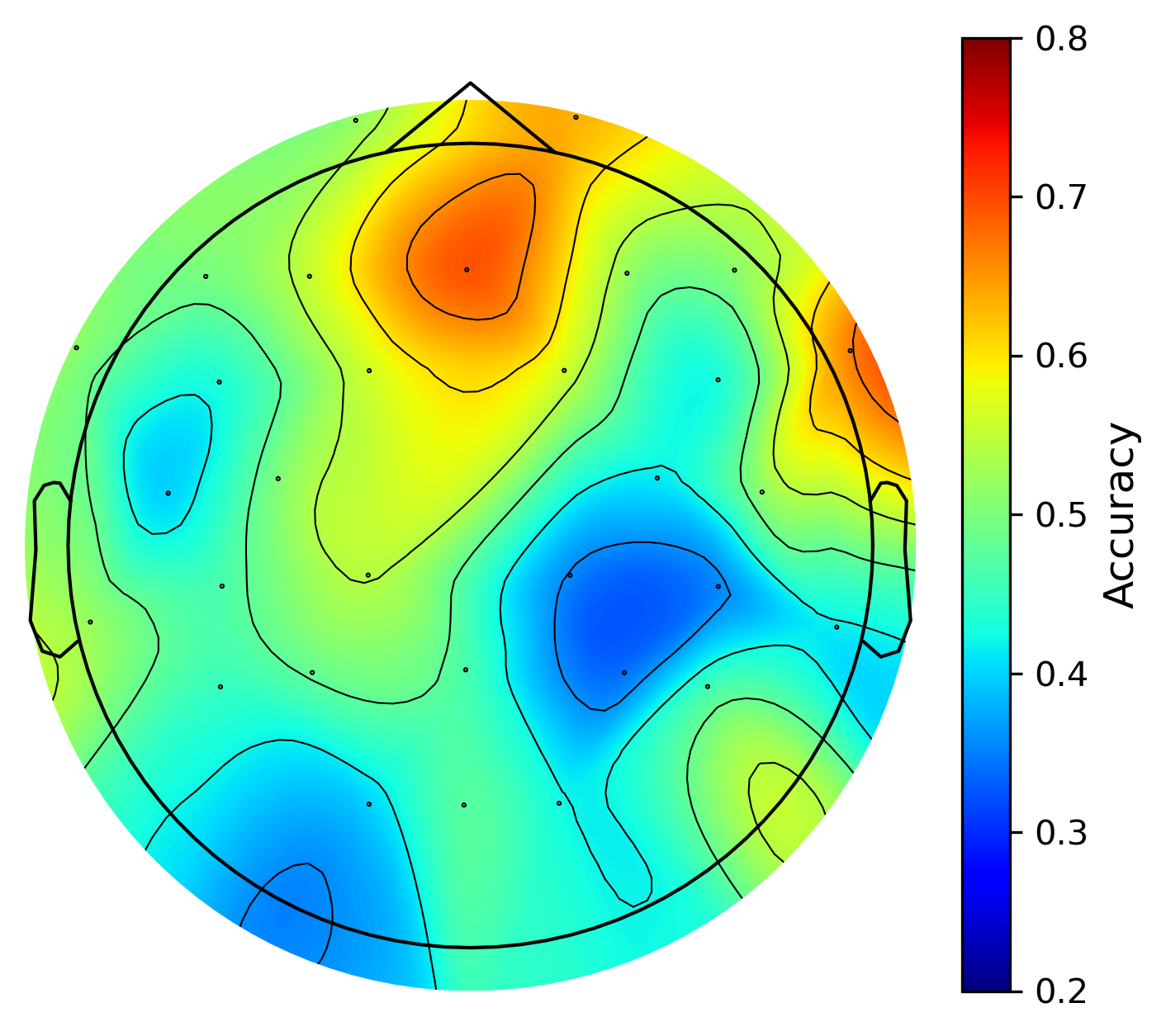}
    }
    \caption{Mean base-models' accuracy per electrode and category comparison (\textcolor{black}{discussed in Section \ref{sssec:between_subject_classification}}).}
    \label{fig:BaseModels}
\end{figure*}

\begin{table*}[t!]
\centering
\caption{\textcolor{black}{Significance tests for Relevance, Likelihood of Purchase, and Diversity across recommendation categories. Values in parentheses are medians. $^*p < .05$, $^{**}p < .01$, $^{***}p < .001$ (Bonferroni-corrected).}}
\setlength{\tabcolsep}{5pt}  
\resizebox{0.8\textwidth}{!}{
\begin{tabular}{|l|l|l|l|l|}
\hline
\textbf{Test} & \textbf{Comparison} & \textbf{Relevance} & \textbf{Likelihood of Purchase} & \textbf{Diversity} \\
\hline
Friedman &  & $\chi^2(3) = 56.37^{***}$ & $\chi^2(3) = 50.74^{***}$ & $\chi^2(3) = 31.67^{***}$ \\
\hline
\multirow{6}{*}{Paired Wilcoxon} 
 & \textbf{E}xact vs \textbf{S}ubstitute & $W = 3^{**}$ (5 vs. 4) & $W = 7.5^{**}$ (4 vs. 4) & $W = 15.5^{**}$ (2 vs. 3) \\
 & \textbf{E}xact vs \textbf{C}omplement & $W = 0^{***}$ (5 vs. 4) & $W = 0^{***}$ (4 vs. 4) & $W = 10^{***}$ (2 vs. 3) \\
 & \textbf{E}xact vs \textbf{I}rrelevant & $W = 0^{***}$ (5 vs. 1) & $W = 0^{***}$ (4 vs. 1) & $W = 13^{***}$ (2 vs. 5) \\
 & \textbf{S}ubstitute vs \textbf{C}omplement & $W = 36.5^{*}$ (4 vs. 4) & $W = 18^{**}$ (3 vs. 3) & $W = 17.5^{**}$ (2 vs. 3) \\
 & \textbf{S}ubstitute vs \textbf{I}rrelevant & $W = 0^{***}$ (4 vs. 1) & $W = 0^{***}$ (3 vs. 1) & $W = 12^{***}$ (2 vs. 5) \\
 & \textbf{C}omplement vs \textbf{I}rrelevant & $W = 0^{***}$ (4 vs. 1) & $W = 12^{***}$ (3 vs. 1) & $W = 14^{**}$ (3 vs. 5) \\
\hline
\end{tabular}}
\label{tab:beh_score_test_results}
\end{table*}

\subsubsection{Between-subject experiments}
The results from the meta-models trained on all subjects show that, for certain comparisons, accuracies are comparable to or even higher than those of individual models after excluding chance-level performers. This suggests that using ERPs may help reduce noise and leverage data from all participants to identify more robust neural patterns. However, for some comparisons, the mean meta-accuracies remain around chance level, indicating that the advantages of this approach may not be uniform across all conditions. The average accuracies of the meta-models for each comparison are presented in \Cref{tab:accuracy_metamodel}.

Consistent with the previous results, the ``E vs. S'' comparison yielded the lowest mean accuracy. In contrast, comparisons involving products relevant to the query (\textbf{E}xact or \textbf{C}omplement) and irrelevant ones remain among the top performers. Unexpectedly, the highest-performing classifier is the ``C vs. S'' comparison, which was associated with one of the lowest performances in the previous section. These results might suggest that the distinction between \textbf{C}omplement and \textbf{S}ubstitute products are linked to subtle neural patterns that become more apparent when working with \glspl{erp}.

The topographic accuracy distribution of channel-based models (\Cref{fig:BaseModels}) shows that the lowest-performing meta-model (``E vs. S'') lacked electrodes with accuracy above 0.6, reinforcing previous findings. In contrast, the best-performing models consistently captured distinct activity patterns in specific brain regions. Notably, variations in the mid and left prefrontal cortex correlated with higher predictive power, suggesting its role in relevance assessment \cite{Liang2017} (``E'', ``C'', and ``S'' vs. ``I'') and concept learning \cite{Mack2020-cn, Liang2017} (``E vs C''). The prefrontal cortex, key to executive functions such as planning, decision-making, and social behavior, integrates information broadly, while its left ventrolateral region is particularly specialized for language and verbal memory \cite{Mack2020,Liang2017}.

Moreover, the ``C vs. S'' comparison shows a pronounced and extensive left-lateralization difference, which may reflect distinct involvement of a range of cognitive tasks. These tasks could include differences in semantic processing (e.g., complement products might be more closely linked to this type of processing), as well as visual attention and top-down modulation \cite{Pantazatos2012-vo} (e.g., substitute products, which are closer to the search, may require more attention to detail). However, the precise nature of this difference remains uncertain, and further investigation is needed to determine its significance.

In summary, neural responses to \textbf{E}xact and \textbf{S}ubstitute product recommendations are largely similar, with greater differentiation for more distinct product types. The ``C vs. S'' comparison ranks among the best-performing models, indicating that subtle neural patterns emerge with \glspl{erp}. Topographic analysis highlights the prefrontal cortex as central to these distinctions. Our findings also suggest that evaluating \textbf{S}ubstitute and \textbf{C}omplement products engages broader cognitive processes.

\subsection{Behavioural variation across recommendation categories}
\label{ssec:rq2_results}

This section addresses \textbf{RQ2:} \textit{How do different recommendation categories influence users' perception (and subsequent observable behaviour) regarding item relevance, purchaseability, and recommendation diversity?} We find that, on average, users correctly classify 71\% of product categories, with most confusion occurring between the \textbf{E}xact and \textbf{S}ubstitute categories. A chi-square test of independence (\(X^2(16) = 9.96, \, p = .86\)) revealed no significant differences between expected and observed confusion matrices, indicating that perceived recommendation categories align with predefined labels.

\begin{figure*}[t!]
    \def\w{0.3\linewidth}
    \def\h{0.2\linewidth}  
    \subfloat[Relevance vs. Likelihood of purchase ($p$=.02, $r$=.68).]
    {
        \includegraphics[trim={0 0 100 0},clip, width=\w, height=\h, keepaspectratio]{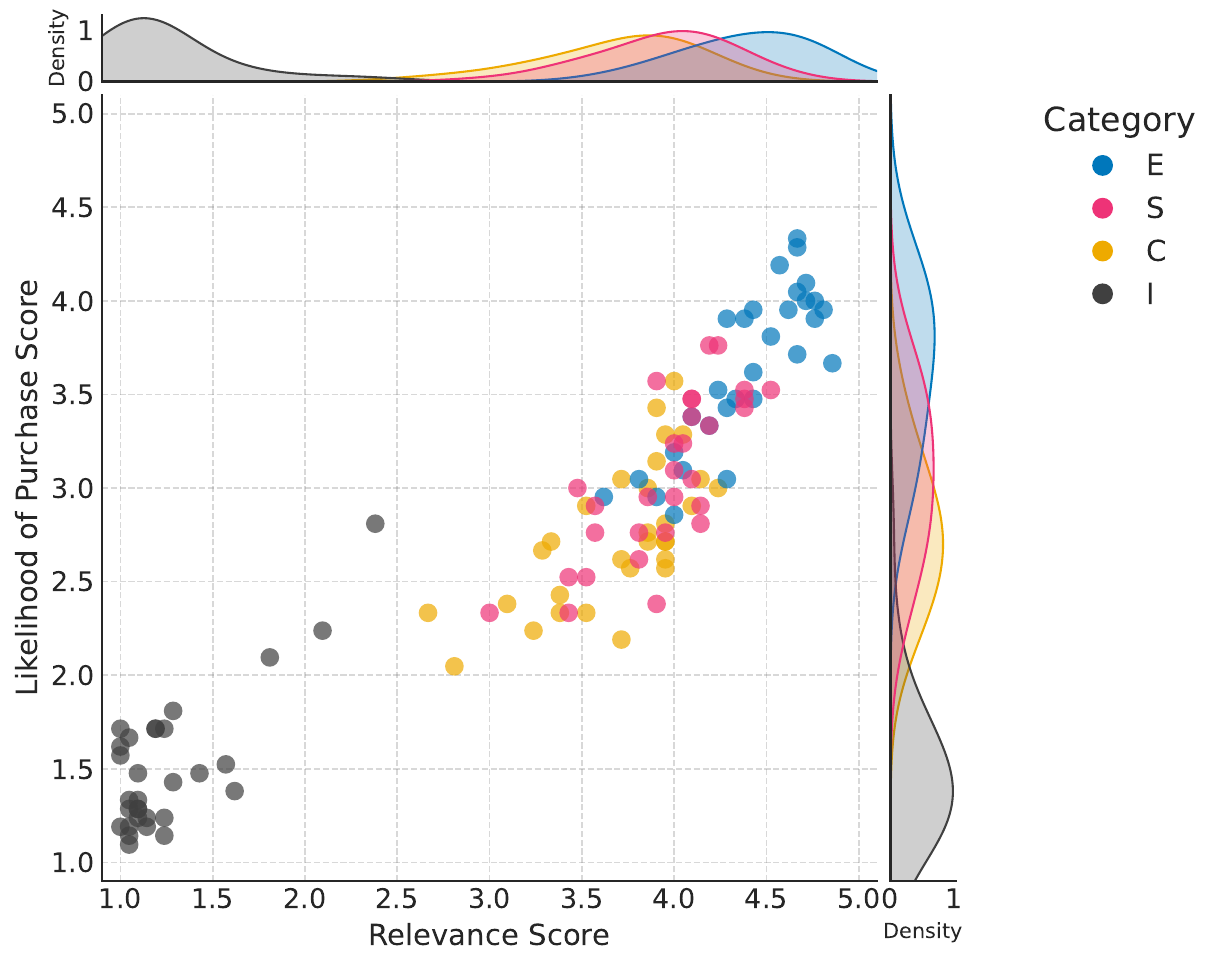}
    }
    \hspace{0.01\linewidth}
    \subfloat[Diversity vs. Relevance ($p$=.01, $r$=-.56).]
    {
        \includegraphics[trim={0 0 100 0},clip, width=\w, height=\h, keepaspectratio]{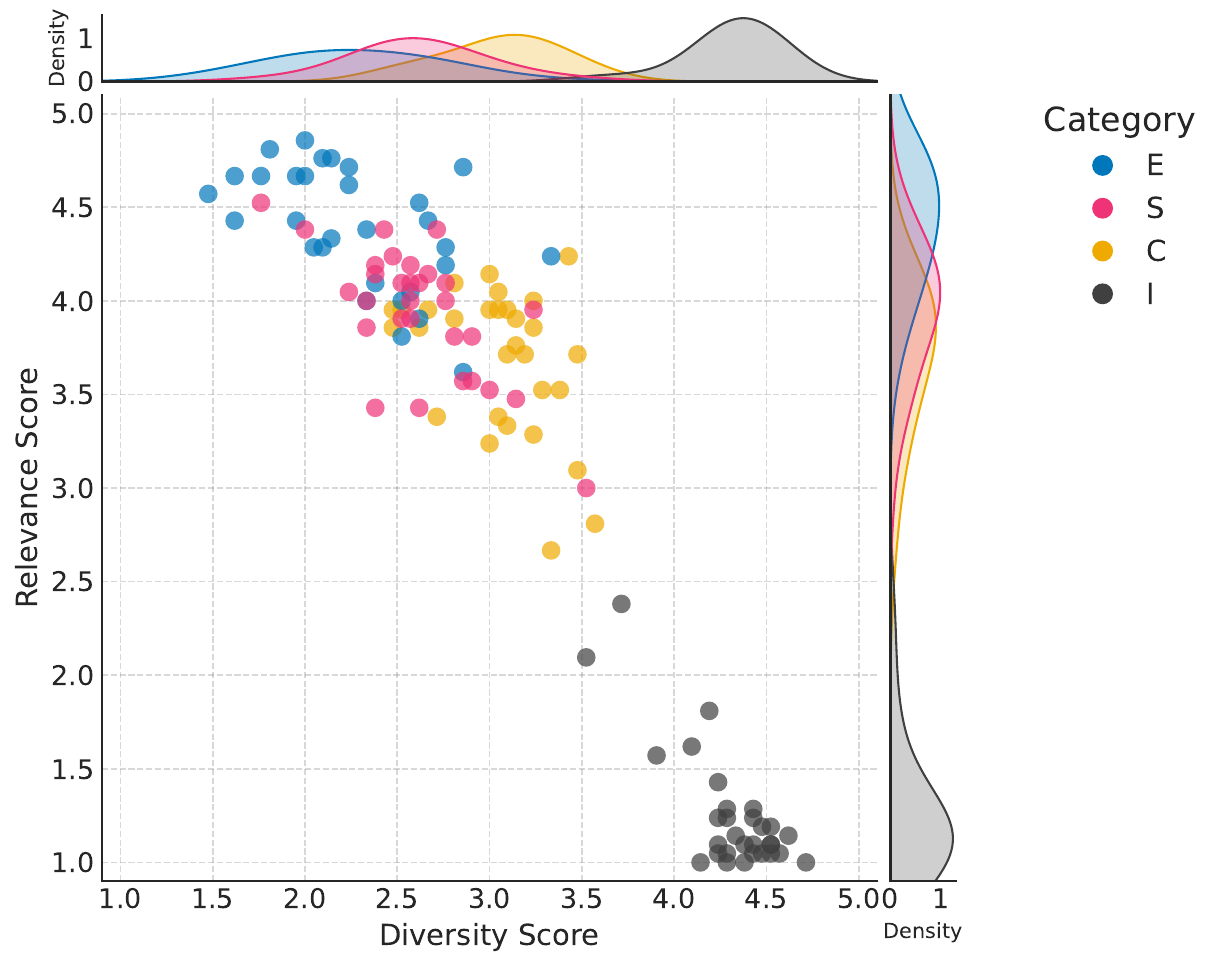}
    }
    \hspace{0.01\linewidth}
    \subfloat[Diversity vs Likelihood of purchase ($p$=.1, $r$=-.42).]
    {
        \includegraphics[width=\w, height=\h, keepaspectratio]{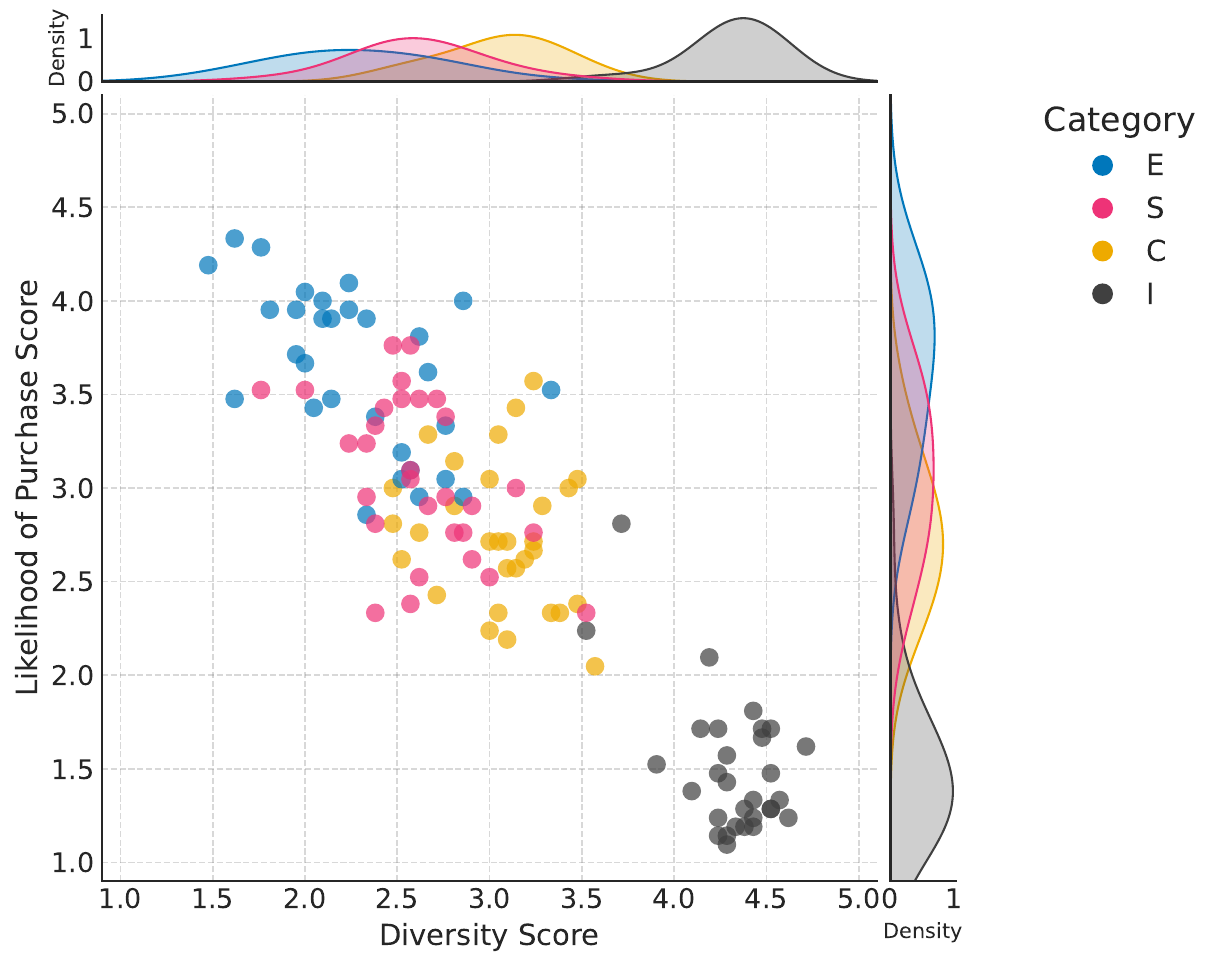}
    }
    \vspace{-5pt}
    \caption{\textcolor{black}{Scatter plots comparing average rating scores including Pearson Correlation values between ratings (each dot represents mean score per stimulus, per category).}}
    \label{fig:scatterplots}
\end{figure*}

Considering all product categories, our analysis ( \Cref{tab:beh_score_test_results} and \Cref{fig:scatterplots}) reveals that \textbf{E}xact and \textbf{I}rrelevant recommendations lie at the extremes across Relevance, Likelihood of Purchase, and Diversity scales. In contrast, \textbf{S}ubstitute and \textbf{C}omplement recommendations score in the mid-range. The Diversity vs. Likelihood and Diversity vs. Relevance plots show that while \textbf{E}xact and \textbf{I}rrelevant recommendations are distinctly separated at the high and low ends, the clusters for \textbf{S}ubstitute and \textbf{C}omplement tend to overlap---\textbf{S}ubstitute with \textbf{E}xact at the upper end and with \textbf{C}omplement at the lower end. The Relevance vs. Likelihood plot indicates that \textbf{E}xact, \textbf{S}ubstitute, and \textbf{C}omplement recommendations achieve above-average scores. This suggests that \textbf{C}omplement and \textbf{S}ubstitute recommendations can improve product diversity without sacrificing relevance.

Furthermore, \Cref{fig:scatterplots} shows correlation trends of the behavioural scales, regardless of categories. As expected, Relevance correlates positively with  \gls{lop} but negatively with Diversity. Interestingly, while \gls{lop} also shows a negative correlation with Diversity, the effect is weaker and not statistically significant. This suggests that intermediate recommendations, such as \textbf{S}ubstitute and \textbf{C}omplement, strike a balance---maintaining relevance while enhancing diversity. Thus, \gls{rs} could improve user experience by incorporating these intermediate categories rather than relying solely on highly relevant or novel recommendations. 

Last, we analysed behaviour across classification types. We applied the Shapiro-Wilk and Levene tests, which revealed significant deviations from normality and homogeneity. Then, we used repeated-measures Friedman and paired Wilcoxon signed-rank tests. Our results demonstrate the importance of designing \gls{rs} that account for behavioural variability (as shown in \Cref{tab:beh_score_test_results}), suggesting that integrating intermediate recommendation types enhance perceived relevance and foster a more diverse, engaging experience.

\subsection{Engagement profiling}
\label{ssec:}

\begin{figure}[t!]
    \begin{center}
    \includegraphics[width=0.9\linewidth]{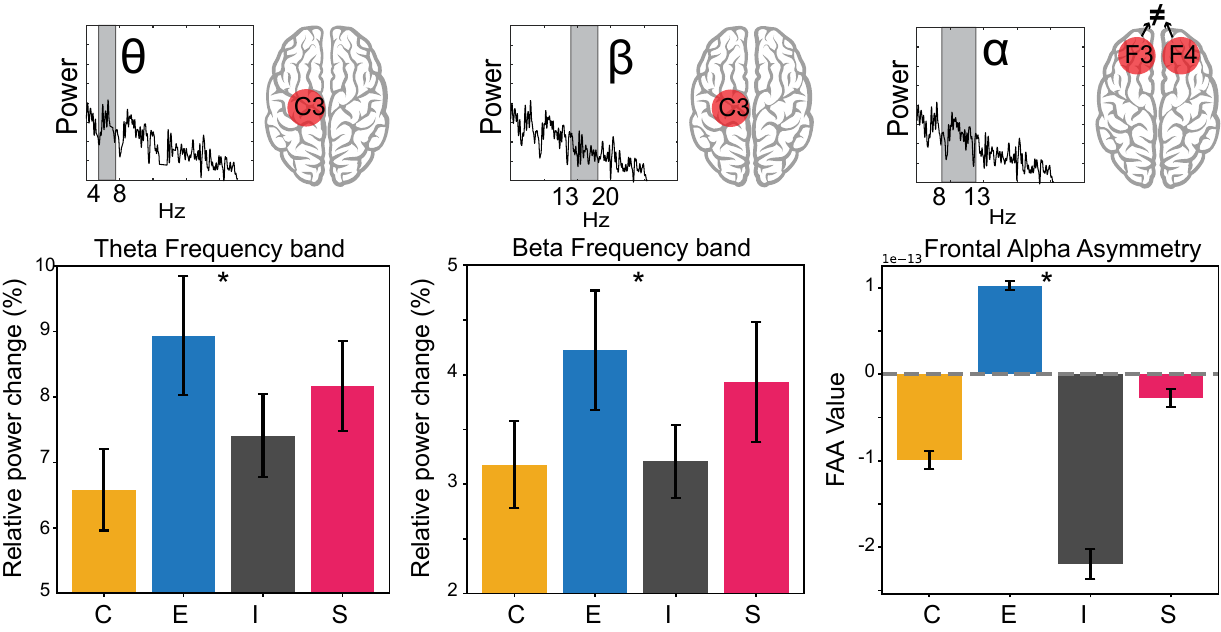}
    \end{center}
    \caption{Engagement across recommendation categories.}
    \label{fig:engagement}
\end{figure}

This section addresses \textbf{RQ3:} \textit{Can we identify neurophysiological markers of engagement that systematically vary across recommendation categories, and what do these variations reveal about the effectiveness of different recommendation strategies?} To investigate differences in engagement metrics (\textcolor{black}{e.g., \gls{faa}, Theta and Beta power; Section \ref{sssec:between_subjects_engagement_analysis})} across recommendation categories, we conducted a repeated measures ANOVA with category (Exact, Irrelevant, Substitute, Complement) 
as the within-subject factor (\Cref{fig:engagement}). For \gls{faa}, we observed a significant main effect of category (\textit{F}(3, 56) = 2.91, \textit{p} < .05), indicating that FAA varied across recommendation categories. The \textbf{E}xact condition showed the highest and only positive mean value (Mean = 9.91e-14). This finding indicates that this condition elicited greater left frontal activity, which has been associated with approach motivation and positive affect. 

Similarly, for Theta power, there was a significant main effect of Category
(\textit{F}(3, 56) = 7.63, \textit{p} < .01). Also, the \textbf{E}xact condition showed to elicit the highest value (Mean = 8.9). We noted the same with Beta power (\textit{F}(3, 56) = 3.5, \textit{p} < .05) where the \textbf{E}xact condition showed the highest value (Mean = 4.22). These findings reinforce that \textbf{E}xact recommendations are associated with more positive cognitive and affective responses, as reflected in neural activity.

\section{Conclusions}
\label{sec:conclusion}

\textcolor{black}{Recommender systems have traditionally prioritised relevance, often neglecting how different types of recommendations shape not only user ratings but also underlying neural processing and decision strategies. Specifically, our study demonstrated that \textbf{E}xact and \textbf{S}ubstitute items evoke comparable cognitive responses and user evaluations, while \textbf{C}omplement recommendations strike a balance between high relevance, purchasability and diversity, outperforming \textbf{I}rrelevant items without sacrificing novelty. By leveraging item‑to‑item relationships from historical interactions, \gls{rs} could improve recommendation quality and even alleviate cold‑start issues for new users or products.}

\textcolor{black}{Moreover, we observed notable individual differences---some users clearly distinguish between recommendation categories (\autoref{fig:acc_plots}; \autoref{ssec:rq2_results}), whereas others react more uniformly, and perceptions of \textbf{C}omplement items range from ``irrelevant clutter'' to ``valuable extras.'' Such variability highlights the shortcomings of one‑size‑fits‑all approaches and points toward multi‑objective, personalised strategies that integrate richer behavioural---and where possible, neural---signals. For instance, neural and behavioural user signals could be utilized to personalize complementary item recommendations in a manner similar to \citet{Yan2022}, offering an advantage through the incorporation of user signals. Looking beyond e‑commerce, these principles of contextualisation and adaptive, user‑specific tuning are likely to improve recommendation effectiveness in other domains like music, video and gaming.}

\textcolor{black}{Despite its insights, this study has some limitations. First, while \gls{eeg} captures neural dynamics well, its low spatial resolution limits source localisation compared to fMRI or MEG. It is also prone to motion and electromagnetic artifacts; although preprocessing was applied, some noise remains. Future work could use more robust artifact rejection or combine \gls{eeg} with other physiological measures. Second, the lab setting reduces confounds but may not reflect real-world cognition and emotion. Ecologically valid setups, e.g., real-time, adaptive interactions, could improve generalisability. Finally, categorisation is subjective and shaped by individual experience; personalised models could help capture this variability.}

\begin{acks}

This work was supported by Horizon Europe's European Innovation Council under Pathfinder project SYMBIOTIK (grant 101071147) and by the Agència de Gestió d’Ajuts Universitaris i de Recerca (AGAUR), Generalitat de Catalunya (grant 2023 DI 00080).

\end{acks}

\bibliographystyle{ACM-Reference-Format}
\bibliography{sample-base}

\end{document}

%% file: glossary.tex
\newacronym{ai}{AI}{Artificial Intelligence}
\newacronym{auc}{AUC}{Area Under the Curve}
\newacronym{dl}{DL}{Deep Learning}
\newacronym{ml}{ML}{Machine Learning}
\newacronym{mlp}{MLP}{Multi-Layer Perceptron}
\newacronym{mse}{MSE}{Mean Squared Error}
\newacronym{sota}{SOTA}{state of the art}
\newacronym{ir}{IR}{Information Retrieval}
\newacronym{rs}{RS}{Recommender Systems}
\newacronym{eeg}{EEG}{Electroencephalography}
\newacronym[plural={ERP's},firstplural={Event Related Potentials}]{erp}{ERP}{Event Related Potential}
\newacronym{kl}{KL}{Kullback-Leibler}
\newacronym{ica}{ICA}{Independent Component Analysis}
\newacronym{svm}{SVM}{Support Vector Machine}
\newacronym{snr}{SNR}{Signal-to-Noise Ratio}
\newacronym{pca}{PCA}{Principal Component Analysis}
\newacronym{dfa}{DFA}{Detrended Fluctuation Analysis}
\newacronym{faa}{FAA}{Frontal Alpha Asymmetry}
\newacronym{ci}{CI}{Choice Index}
\newacronym{psd}{PSD}{Power Spectral Density}
\newacronym{bci}{BCI}{Brain-Computer Interface}
\newacronym{llm}{LLMs}{Large Language Models}

\newacronym{lop}{LoP}{Likelihood of Purchase}

%% file: main.bbl

\begin{thebibliography}{76}


\ifx \showCODEN    \undefined \def \showCODEN     #1{\unskip}     \fi
\ifx \showDOI      \undefined \def \showDOI       #1{#1}\fi
\ifx \showISBNx    \undefined \def \showISBNx     #1{\unskip}     \fi
\ifx \showISBNxiii \undefined \def \showISBNxiii  #1{\unskip}     \fi
\ifx \showISSN     \undefined \def \showISSN      #1{\unskip}     \fi
\ifx \showLCCN     \undefined \def \showLCCN      #1{\unskip}     \fi
\ifx \shownote     \undefined \def \shownote      #1{#1}          \fi
\ifx \showarticletitle \undefined \def \showarticletitle #1{#1}   \fi
\ifx \showURL      \undefined \def \showURL       {\relax}        \fi
\providecommand\bibfield[2]{#2}
\providecommand\bibinfo[2]{#2}
\providecommand\natexlab[1]{#1}
\providecommand\showeprint[2][]{arXiv:#2}

\bibitem[Abdulkarim and Al-Faiz(2021)]%
        {Abdulkarim2021}
\bibfield{author}{\bibinfo{person}{Haider Abdulkarim} {and}
  \bibinfo{person}{Mohammed Al-Faiz}.} \bibinfo{year}{2021}\natexlab{}.
\newblock \showarticletitle{Online multiclass EEG feature extraction and
  recognition using modified convolutional neural network method}.
\newblock \bibinfo{journal}{\emph{International Journal of Electrical and
  Computer Engineering}}  \bibinfo{volume}{11} (\bibinfo{date}{05}
  \bibinfo{year}{2021}), \bibinfo{pages}{4016--4026}.
\newblock
\urldef\tempurl%
\url{https://doi.org/10.11591/ijece.v11i5.pp4016-4026}
\showDOI{\tempurl}


\bibitem[Adomavicius and Tuzhilin(2005)]%
        {adomavicius}
\bibfield{author}{\bibinfo{person}{Gediminas Adomavicius} {and}
  \bibinfo{person}{Alexander Tuzhilin}.} \bibinfo{year}{2005}\natexlab{}.
\newblock \showarticletitle{Toward the next generation of recommender systems:
  A survey of the state-of-the-art and possible extensions}.
\newblock \bibinfo{journal}{\emph{Knowledge and Data Engineering, IEEE
  Transactions on}}  \bibinfo{volume}{17} (\bibinfo{date}{07}
  \bibinfo{year}{2005}), \bibinfo{pages}{734--749}.
\newblock
\urldef\tempurl%
\url{https://doi.org/10.1109/TKDE.2005.99}
\showDOI{\tempurl}


\bibitem[Aldayel et~al\mbox{.}(2020)]%
        {aldayel2020deep}
\bibfield{author}{\bibinfo{person}{Mashael Aldayel}, \bibinfo{person}{Mourad
  Ykhlef}, {and} \bibinfo{person}{Abeer Al-Nafjan}.}
  \bibinfo{year}{2020}\natexlab{}.
\newblock \showarticletitle{Deep learning for EEG-based preference
  classification in neuromarketing}.
\newblock \bibinfo{journal}{\emph{Applied Sciences}} \bibinfo{volume}{10},
  \bibinfo{number}{4} (\bibinfo{year}{2020}), \bibinfo{pages}{1525}.
\newblock


\bibitem[Anderson et~al\mbox{.}(2020)]%
        {Anderson2020}
\bibfield{author}{\bibinfo{person}{Ashton Anderson}, \bibinfo{person}{Lucas
  Maystre}, \bibinfo{person}{Ian Anderson}, \bibinfo{person}{Rishabh Mehrotra},
  {and} \bibinfo{person}{Mounia Lalmas}.} \bibinfo{year}{2020}\natexlab{}.
\newblock \showarticletitle{Algorithmic Effects on the Diversity of Consumption
  on Spotify}. In \bibinfo{booktitle}{\emph{Proceedings of The Web Conference
  2020}} (Taipei, Taiwan) \emph{(\bibinfo{series}{WWW '20})}.
  \bibinfo{publisher}{Association for Computing Machinery},
  \bibinfo{address}{New York, NY, USA}, \bibinfo{pages}{2155–2165}.
\newblock
\showISBNx{9781450370233}
\urldef\tempurl%
\url{https://doi.org/10.1145/3366423.3380281}
\showDOI{\tempurl}


\bibitem[Antognini et~al\mbox{.}(2021)]%
        {ijcai2021p72}
\bibfield{author}{\bibinfo{person}{Diego Antognini}, \bibinfo{person}{Claudiu
  Musat}, {and} \bibinfo{person}{Boi Faltings}.}
  \bibinfo{year}{2021}\natexlab{}.
\newblock \showarticletitle{Interacting with Explanations through Critiquing}.
  In \bibinfo{booktitle}{\emph{Proceedings of the Thirtieth International Joint
  Conference on Artificial Intelligence, {IJCAI-21}}},
  \bibfield{editor}{\bibinfo{person}{Zhi-Hua Zhou}} (Ed.).
  \bibinfo{publisher}{International Joint Conferences on Artificial
  Intelligence Organization}, \bibinfo{pages}{515--521}.
\newblock
\urldef\tempurl%
\url{https://doi.org/10.24963/ijcai.2021/72}
\showDOI{\tempurl}
\newblock
\shownote{Main Track}.


\bibitem[Antognini et~al\mbox{.}(2020)]%
        {Antognini2020TRECSAT}
\bibfield{author}{\bibinfo{person}{Diego Antognini},
  \bibinfo{person}{Claudiu~Cristian Musat}, {and} \bibinfo{person}{Boi
  Faltings}.} \bibinfo{year}{2020}\natexlab{}.
\newblock \showarticletitle{T-RECS: a Transformer-based Recommender Generating
  Textual Explanations and Integrating Unsupervised Language-based Critiquing}.
\newblock \bibinfo{journal}{\emph{ArXiv}}  \bibinfo{volume}{abs/2005.11067}
  (\bibinfo{year}{2020}).
\newblock
\urldef\tempurl%
\url{https://api.semanticscholar.org/CorpusID:218863076}
\showURL{%
\tempurl}


\bibitem[Bandara et~al\mbox{.}(2021)]%
        {4}
\bibfield{author}{\bibinfo{person}{Sathsarani~K. Bandara},
  \bibinfo{person}{Uvini~C. Wijesinghe}, \bibinfo{person}{Badra~P. Jayalath},
  \bibinfo{person}{Saumya~K. Bandara}, \bibinfo{person}{Prasanna~S. Haddela},
  {and} \bibinfo{person}{Lumini~M. Wickramasinghe}.}
  \bibinfo{year}{2021}\natexlab{}.
\newblock \showarticletitle{EEG Based Neuromarketing Recommender System for
  Video Commercials}. In \bibinfo{booktitle}{\emph{2021 IEEE 16th International
  Conference on Industrial and Information Systems (ICIIS)}}.
\newblock
\urldef\tempurl%
\url{https://doi.org/10.1109/ICIIS53135.2021.9660742}
\showDOI{\tempurl}


\bibitem[Barbe et~al\mbox{.}(2009)]%
        {barbe2009welch}
\bibfield{author}{\bibinfo{person}{Kurt Barbe}, \bibinfo{person}{Rik Pintelon},
  {and} \bibinfo{person}{Johan Schoukens}.} \bibinfo{year}{2009}\natexlab{}.
\newblock \showarticletitle{Welch method revisited: nonparametric power
  spectrum estimation via circular overlap}.
\newblock \bibinfo{journal}{\emph{IEEE Transactions on signal processing}}
  \bibinfo{volume}{58}, \bibinfo{number}{2} (\bibinfo{year}{2009}),
  \bibinfo{pages}{553--565}.
\newblock


\bibitem[Barros et~al\mbox{.}(2022)]%
        {barros2022frontal}
\bibfield{author}{\bibinfo{person}{Catarina Barros}, \bibinfo{person}{Ana~Rita
  Pereira}, \bibinfo{person}{Adriana Sampaio}, \bibinfo{person}{Ana Buj{\'a}n},
  {and} \bibinfo{person}{Diego Pinal}.} \bibinfo{year}{2022}\natexlab{}.
\newblock \showarticletitle{Frontal alpha asymmetry and negative Mood: a
  cross-sectional study in older and younger adults}.
\newblock \bibinfo{journal}{\emph{Symmetry}} \bibinfo{volume}{14},
  \bibinfo{number}{8} (\bibinfo{year}{2022}), \bibinfo{pages}{1579}.
\newblock


\bibitem[Buzsaki(2006)]%
        {buzsaki2006rhythms}
\bibfield{author}{\bibinfo{person}{Gyorgy Buzsaki}.}
  \bibinfo{year}{2006}\natexlab{}.
\newblock \bibinfo{booktitle}{\emph{Rhythms of the Brain}}.
\newblock \bibinfo{publisher}{Oxford university press}.
\newblock


\bibitem[Buzs{\'a}ki and da~Silva(2012)]%
        {buzsaki2012high}
\bibfield{author}{\bibinfo{person}{Gy{\"o}rgy Buzs{\'a}ki} {and}
  \bibinfo{person}{Fernando~Lopes da Silva}.} \bibinfo{year}{2012}\natexlab{}.
\newblock \showarticletitle{High frequency oscillations in the intact brain}.
\newblock \bibinfo{journal}{\emph{Progress in neurobiology}}
  \bibinfo{volume}{98}, \bibinfo{number}{3} (\bibinfo{year}{2012}),
  \bibinfo{pages}{241--249}.
\newblock


\bibitem[Claypool et~al\mbox{.}(2001)]%
        {claypool2001implicit}
\bibfield{author}{\bibinfo{person}{Mark Claypool}, \bibinfo{person}{Phong Le},
  \bibinfo{person}{Makoto Wased}, {and} \bibinfo{person}{David Brown}.}
  \bibinfo{year}{2001}\natexlab{}.
\newblock \showarticletitle{Implicit interest indicators}. In
  \bibinfo{booktitle}{\emph{Proceedings of the 6th international conference on
  Intelligent user interfaces}}. \bibinfo{pages}{33--40}.
\newblock


\bibitem[Da'u and Salim(2020)]%
        {93.2}
\bibfield{author}{\bibinfo{person}{Aminu Da'u} {and} \bibinfo{person}{Naomie
  Salim}.} \bibinfo{year}{2020}\natexlab{}.
\newblock \showarticletitle{Recommendation system based on deep learning
  methods: a systematic review and new directions}.
\newblock \bibinfo{journal}{\emph{Artificial Intelligence Review}}
  \bibinfo{volume}{53}, \bibinfo{number}{4} (\bibinfo{year}{2020}),
  \bibinfo{pages}{2709--2748}.
\newblock
\urldef\tempurl%
\url{https://doi.org/10.1007/s10462-019-09744-1}
\showDOI{\tempurl}


\bibitem[Davis et~al\mbox{.}(2023)]%
        {RS_16}
\bibfield{author}{\bibinfo{person}{Keith~M. Davis}, \bibinfo{person}{Michiel
  Spape}, {and} \bibinfo{person}{Tuukka Ruotsalo}.}
  \bibinfo{year}{2023}\natexlab{}.
\newblock \showarticletitle{Contradicted by the Brain: Predicting Individual
  and Group Preferences via Brain-Computer Interfacing}.
\newblock \bibinfo{journal}{\emph{IEEE Transactions on Affective Computing}}
  \bibinfo{volume}{14}, \bibinfo{number}{4} (\bibinfo{year}{2023}),
  \bibinfo{pages}{3094--3105}.
\newblock
\urldef\tempurl%
\url{https://doi.org/10.1109/TAFFC.2022.3225885}
\showDOI{\tempurl}


\bibitem[Deldjoo et~al\mbox{.}(2024)]%
        {deldjoo2024reviewmodernrecommendersystems}
\bibfield{author}{\bibinfo{person}{Yashar Deldjoo}, \bibinfo{person}{Zhankui
  He}, \bibinfo{person}{Julian McAuley}, \bibinfo{person}{Anton Korikov},
  \bibinfo{person}{Scott Sanner}, \bibinfo{person}{Arnau Ramisa},
  \bibinfo{person}{René Vidal}, \bibinfo{person}{Maheswaran Sathiamoorthy},
  \bibinfo{person}{Atoosa Kasirzadeh}, {and} \bibinfo{person}{Silvia Milano}.}
  \bibinfo{year}{2024}\natexlab{}.
\newblock \bibinfo{title}{A Review of Modern Recommender Systems Using
  Generative Models (Gen-RecSys)}.
\newblock
\newblock
\showeprint[arxiv]{2404.00579}~[cs.IR]
\urldef\tempurl%
\url{https://arxiv.org/abs/2404.00579}
\showURL{%
\tempurl}


\bibitem[Esteller et~al\mbox{.}(2001)]%
        {904882}
\bibfield{author}{\bibinfo{person}{R. Esteller}, \bibinfo{person}{G.
  Vachtsevanos}, \bibinfo{person}{J. Echauz}, {and} \bibinfo{person}{B. Litt}.}
  \bibinfo{year}{2001}\natexlab{}.
\newblock \showarticletitle{A comparison of waveform fractal dimension
  algorithms}.
\newblock \bibinfo{journal}{\emph{IEEE Transactions on Circuits and Systems I:
  Fundamental Theory and Applications}} \bibinfo{volume}{48},
  \bibinfo{number}{2} (\bibinfo{year}{2001}), \bibinfo{pages}{177--183}.
\newblock
\urldef\tempurl%
\url{https://doi.org/10.1109/81.904882}
\showDOI{\tempurl}


\bibitem[Eugster et~al\mbox{.}(2014)]%
        {Tukka}
\bibfield{author}{\bibinfo{person}{Manuel Eugster}, \bibinfo{person}{Tuukka
  Ruotsalo}, \bibinfo{person}{Michiel Spapé}, \bibinfo{person}{Ilkka Kosunen},
  \bibinfo{person}{Oswald Barral}, \bibinfo{person}{Niklas Ravaja},
  \bibinfo{person}{Giulio Jacucci}, {and} \bibinfo{person}{Samuel Kaski}.}
  \bibinfo{year}{2014}\natexlab{}.
\newblock \showarticletitle{Predicting term-relevance from brain signals}.
\newblock  (\bibinfo{date}{07} \bibinfo{year}{2014}).
\newblock
\showISBNx{978-1-4503-2257-7}
\urldef\tempurl%
\url{https://doi.org/10.1145/2600428.2609594}
\showDOI{\tempurl}


\bibitem[Gao(2022)]%
        {98}
\bibfield{author}{\bibinfo{person}{Chen Gao}.} \bibinfo{year}{2022}\natexlab{}.
\newblock \showarticletitle{Causal Inference in Recommender Systems: A Survey
  and Future Directions}.
\newblock \bibinfo{journal}{\emph{arXiv preprint arXiv:2208.12397}}
  (\bibinfo{date}{Dec} \bibinfo{year}{2022}).
\newblock
\newblock
\shownote{Version 2}.


\bibitem[Gotlib(1998)]%
        {gotlib1998eeg}
\bibfield{author}{\bibinfo{person}{Ian~H Gotlib}.}
  \bibinfo{year}{1998}\natexlab{}.
\newblock \showarticletitle{EEG alpha asymmetry, depression, and cognitive
  functioning}.
\newblock \bibinfo{journal}{\emph{Cognition \& Emotion}} \bibinfo{volume}{12},
  \bibinfo{number}{3} (\bibinfo{year}{1998}), \bibinfo{pages}{449--478}.
\newblock


\bibitem[Gramfort et~al\mbox{.}(2013)]%
        {GramfortEtAl2013a}
\bibfield{author}{\bibinfo{person}{Alexandre Gramfort}, \bibinfo{person}{Martin
  Luessi}, \bibinfo{person}{Eric Larson}, \bibinfo{person}{Denis~A. Engemann},
  \bibinfo{person}{Daniel Strohmeier}, \bibinfo{person}{Christian Brodbeck},
  \bibinfo{person}{Roman Goj}, \bibinfo{person}{Mainak Jas},
  \bibinfo{person}{Teon Brooks}, \bibinfo{person}{Lauri Parkkonen}, {and}
  \bibinfo{person}{Matti~S. H{\"a}m{\"a}l{\"a}inen}.}
  \bibinfo{year}{2013}\natexlab{}.
\newblock \showarticletitle{{{MEG}} and {{EEG}} Data Analysis with
  {{MNE}}-{{Python}}}.
\newblock \bibinfo{journal}{\emph{Frontiers in Neuroscience}}
  \bibinfo{volume}{7}, \bibinfo{number}{267} (\bibinfo{year}{2013}),
  \bibinfo{pages}{1--13}.
\newblock
\urldef\tempurl%
\url{https://doi.org/10.3389/fnins.2013.00267}
\showDOI{\tempurl}


\bibitem[Gupta et~al\mbox{.}(2009)]%
        {eegKl}
\bibfield{author}{\bibinfo{person}{Anjum Gupta}, \bibinfo{person}{Shibin
  Parameswaran}, {and} \bibinfo{person}{Cheng-Han Lee}.}
  \bibinfo{year}{2009}\natexlab{}.
\newblock \showarticletitle{Classification of electroencephalography (EEG)
  signals for different mental activities using Kullback Leibler (KL)
  divergence}, In \bibinfo{booktitle}{Proceedings of the {IEEE} International
  Conference on Acoustics, Speech, and Signal Processing, {ICASSP} 2009, 19-24
  April 2009, Taipei, Taiwan}.
\newblock \bibinfo{journal}{\emph{IEEE Int. Conf. Acoust. Speech Signal
  Process}}, \bibinfo{pages}{1697--1700}.
\newblock
\urldef\tempurl%
\url{https://doi.org/10.1109/ICASSP.2009.4959929}
\showDOI{\tempurl}


\bibitem[Hada et~al\mbox{.}(2021)]%
        {10.1145/3404835.3462939}
\bibfield{author}{\bibinfo{person}{Deepesh~V. Hada},
  \bibinfo{person}{Vijaikumar M.}, {and} \bibinfo{person}{Shirish~K. Shevade}.}
  \bibinfo{year}{2021}\natexlab{}.
\newblock \showarticletitle{ReXPlug: Explainable Recommendation using
  Plug-and-Play Language Model}. In \bibinfo{booktitle}{\emph{Proceedings of
  the 44th International ACM SIGIR Conference on Research and Development in
  Information Retrieval}} (Virtual Event, Canada) \emph{(\bibinfo{series}{SIGIR
  '21})}. \bibinfo{publisher}{Association for Computing Machinery},
  \bibinfo{address}{New York, NY, USA}, \bibinfo{pages}{81–91}.
\newblock
\showISBNx{9781450380379}
\urldef\tempurl%
\url{https://doi.org/10.1145/3404835.3462939}
\showDOI{\tempurl}


\bibitem[Halderman et~al\mbox{.}(2021)]%
        {halderman2021eeg}
\bibfield{author}{\bibinfo{person}{Laura~K Halderman}, \bibinfo{person}{Bridgid
  Finn}, \bibinfo{person}{JR Lockwood}, \bibinfo{person}{Nicole~M Long}, {and}
  \bibinfo{person}{Michael~J Kahana}.} \bibinfo{year}{2021}\natexlab{}.
\newblock \showarticletitle{EEG correlates of engagement during assessment}.
\newblock \bibinfo{journal}{\emph{ETS Research Report Series}}
  \bibinfo{volume}{2021}, \bibinfo{number}{1} (\bibinfo{year}{2021}),
  \bibinfo{pages}{1--17}.
\newblock


\bibitem[Harmon-Jones et~al\mbox{.}(2010)]%
        {harmon2010role}
\bibfield{author}{\bibinfo{person}{Eddie Harmon-Jones},
  \bibinfo{person}{Philip~A Gable}, {and} \bibinfo{person}{Carly~K Peterson}.}
  \bibinfo{year}{2010}\natexlab{}.
\newblock \showarticletitle{The role of asymmetric frontal cortical activity in
  emotion-related phenomena: A review and update}.
\newblock \bibinfo{journal}{\emph{Biological psychology}} \bibinfo{volume}{84},
  \bibinfo{number}{3} (\bibinfo{year}{2010}), \bibinfo{pages}{451--462}.
\newblock


\bibitem[Hu et~al\mbox{.}(2018)]%
        {hu2018reinforcement}
\bibfield{author}{\bibinfo{person}{Yujing Hu}, \bibinfo{person}{Qing Da},
  \bibinfo{person}{Anxiang Zeng}, \bibinfo{person}{Yang Yu}, {and}
  \bibinfo{person}{Yinghui Xu}.} \bibinfo{year}{2018}\natexlab{}.
\newblock \showarticletitle{Reinforcement learning to rank in e-commerce search
  engine: Formalization, analysis, and application}. In
  \bibinfo{booktitle}{\emph{Proceedings of the 24th ACM SIGKDD International
  Conference on Knowledge Discovery \& Data Mining}}.
  \bibinfo{pages}{368--377}.
\newblock


\bibitem[Hu et~al\mbox{.}(2008)]%
        {Hu2008}
\bibfield{author}{\bibinfo{person}{Yifan Hu}, \bibinfo{person}{Yehuda Koren},
  {and} \bibinfo{person}{Chris Volinsky}.} \bibinfo{year}{2008}\natexlab{}.
\newblock \showarticletitle{Collaborative Filtering for Implicit Feedback
  Datasets}. In \bibinfo{booktitle}{\emph{2008 Eighth IEEE International
  Conference on Data Mining}}.
\newblock
\urldef\tempurl%
\url{https://doi.org/10.1109/ICDM.2008.22}
\showDOI{\tempurl}


\bibitem[Ibáñez-Molina and Iglesias-Parro(2014)]%
        {IBANEZMOLINA201469}
\bibfield{author}{\bibinfo{person}{A.J. Ibáñez-Molina} {and}
  \bibinfo{person}{S. Iglesias-Parro}.} \bibinfo{year}{2014}\natexlab{}.
\newblock \showarticletitle{Fractal characterization of internally and
  externally generated conscious experiences}.
\newblock \bibinfo{journal}{\emph{Brain and Cognition}}  \bibinfo{volume}{87}
  (\bibinfo{year}{2014}), \bibinfo{pages}{69--75}.
\newblock
\showISSN{0278-2626}
\urldef\tempurl%
\url{https://doi.org/10.1016/j.bandc.2014.03.002}
\showDOI{\tempurl}


\bibitem[Iyengar and Lepper(1999)]%
        {iyengar1999rethinking}
\bibfield{author}{\bibinfo{person}{Sheena~S Iyengar} {and}
  \bibinfo{person}{Mark~R Lepper}.} \bibinfo{year}{1999}\natexlab{}.
\newblock \showarticletitle{Rethinking the value of choice: a cultural
  perspective on intrinsic motivation.}
\newblock \bibinfo{journal}{\emph{Journal of personality and social
  psychology}} \bibinfo{volume}{76}, \bibinfo{number}{3}
  (\bibinfo{year}{1999}), \bibinfo{pages}{349}.
\newblock


\bibitem[Jacucci et~al\mbox{.}(2019)]%
        {Jacucci2019}
\bibfield{author}{\bibinfo{person}{Giulio Jacucci}, \bibinfo{person}{Oswald
  Barral}, \bibinfo{person}{Pedram Daee}, \bibinfo{person}{Markus Wendt},
  \bibinfo{person}{Baris Serim}, \bibinfo{person}{Tuukka Ruotsalo},
  \bibinfo{person}{Patrik Pluchino}, \bibinfo{person}{Jonathan Freeman},
  \bibinfo{person}{Luciano Gamberini}, \bibinfo{person}{Samuel Kaski}, {and}
  \bibinfo{person}{Benjamin Blankertz}.} \bibinfo{year}{2019}\natexlab{}.
\newblock \showarticletitle{Integrating neurophysiologic relevance feedback in
  intent modeling for information retrieval}.
\newblock \bibinfo{journal}{\emph{Journal of the Association for Information
  Science and Technology}} \bibinfo{volume}{70}, \bibinfo{number}{9}
  (\bibinfo{year}{2019}), \bibinfo{pages}{917--930}.
\newblock
\urldef\tempurl%
\url{https://doi.org/10.1002/asi.24161}
\showDOI{\tempurl}


\bibitem[Jawaheer et~al\mbox{.}(2010)]%
        {jawaheer2010comparison}
\bibfield{author}{\bibinfo{person}{Gawesh Jawaheer}, \bibinfo{person}{Martin
  Szomszor}, {and} \bibinfo{person}{Patty Kostkova}.}
  \bibinfo{year}{2010}\natexlab{}.
\newblock \showarticletitle{Comparison of implicit and explicit feedback from
  an online music recommendation service}. In
  \bibinfo{booktitle}{\emph{proceedings of the 1st international workshop on
  information heterogeneity and fusion in recommender systems}}.
  \bibinfo{pages}{47--51}.
\newblock


\bibitem[Jawaheer et~al\mbox{.}(2014)]%
        {jawaheer2014modeling}
\bibfield{author}{\bibinfo{person}{Gawesh Jawaheer}, \bibinfo{person}{Peter
  Weller}, {and} \bibinfo{person}{Patty Kostkova}.}
  \bibinfo{year}{2014}\natexlab{}.
\newblock \showarticletitle{Modeling user preferences in recommender systems: A
  classification framework for explicit and implicit user feedback}.
\newblock \bibinfo{journal}{\emph{ACM Transactions on Interactive Intelligent
  Systems (TiiS)}} \bibinfo{volume}{4}, \bibinfo{number}{2}
  (\bibinfo{year}{2014}), \bibinfo{pages}{1--26}.
\newblock


\bibitem[Ko et~al\mbox{.}(2022)]%
        {95}
\bibfield{author}{\bibinfo{person}{Hyeyoung Ko}, \bibinfo{person}{Suyeon Lee},
  \bibinfo{person}{Yoonseo Park}, {and} \bibinfo{person}{Anna Choi}.}
  \bibinfo{year}{2022}\natexlab{}.
\newblock \showarticletitle{A Survey of Recommendation Systems: Recommendation
  Models, Techniques, and Application Fields}.
\newblock \bibinfo{journal}{\emph{Electronics}} \bibinfo{volume}{11},
  \bibinfo{number}{1} (\bibinfo{year}{2022}).
\newblock
\showISSN{2079-9292}
\urldef\tempurl%
\url{https://doi.org/10.3390/electronics11010141}
\showDOI{\tempurl}


\bibitem[Lerche(2016)]%
        {lerche2016using}
\bibfield{author}{\bibinfo{person}{Lukas Lerche}.}
  \bibinfo{year}{2016}\natexlab{}.
\newblock \showarticletitle{Using implicit feedback for recommender systems:
  characteristics, applications, and challenges}.
\newblock  (\bibinfo{year}{2016}).
\newblock


\bibitem[Li et~al\mbox{.}(2023a)]%
        {10.1145/3580305.3599535}
\bibfield{author}{\bibinfo{person}{Jiacheng Li}, \bibinfo{person}{Zhankui He},
  \bibinfo{person}{Jingbo Shang}, {and} \bibinfo{person}{Julian McAuley}.}
  \bibinfo{year}{2023}\natexlab{a}.
\newblock \showarticletitle{UCEpic: Unifying Aspect Planning and Lexical
  Constraints for Generating Explanations in Recommendation}. In
  \bibinfo{booktitle}{\emph{Proceedings of the 29th ACM SIGKDD Conference on
  Knowledge Discovery and Data Mining}} (Long Beach, CA, USA)
  \emph{(\bibinfo{series}{KDD '23})}. \bibinfo{publisher}{Association for
  Computing Machinery}, \bibinfo{address}{New York, NY, USA},
  \bibinfo{pages}{1248–1257}.
\newblock
\showISBNx{9798400701030}
\urldef\tempurl%
\url{https://doi.org/10.1145/3580305.3599535}
\showDOI{\tempurl}


\bibitem[Li et~al\mbox{.}(2021)]%
        {Li2021PersonalizedTF}
\bibfield{author}{\bibinfo{person}{Lei Li}, \bibinfo{person}{Yongfeng Zhang},
  {and} \bibinfo{person}{Li Chen}.} \bibinfo{year}{2021}\natexlab{}.
\newblock \showarticletitle{Personalized Transformer for Explainable
  Recommendation}.
\newblock \bibinfo{journal}{\emph{ArXiv}}  \bibinfo{volume}{abs/2105.11601}
  (\bibinfo{year}{2021}).
\newblock
\urldef\tempurl%
\url{https://api.semanticscholar.org/CorpusID:235187032}
\showURL{%
\tempurl}


\bibitem[Li et~al\mbox{.}(2023b)]%
        {10.1145/3580488}
\bibfield{author}{\bibinfo{person}{Lei Li}, \bibinfo{person}{Yongfeng Zhang},
  {and} \bibinfo{person}{Li Chen}.} \bibinfo{year}{2023}\natexlab{b}.
\newblock \showarticletitle{Personalized Prompt Learning for Explainable
  Recommendation}.
\newblock \bibinfo{journal}{\emph{ACM Trans. Inf. Syst.}} \bibinfo{volume}{41},
  \bibinfo{number}{4}, Article \bibinfo{articleno}{103} (\bibinfo{date}{March}
  \bibinfo{year}{2023}), \bibinfo{numpages}{26}~pages.
\newblock
\showISSN{1046-8188}
\urldef\tempurl%
\url{https://doi.org/10.1145/3580488}
\showDOI{\tempurl}


\bibitem[Li and Liu(2022)]%
        {3}
\bibfield{author}{\bibinfo{person}{Sukun Li} {and} \bibinfo{person}{Xiaoxing
  Liu}.} \bibinfo{year}{2022}\natexlab{}.
\newblock \showarticletitle{Toward a BCI-Based Personalized Recommender System
  Using Deep Learning}. In \bibinfo{booktitle}{\emph{2022 IEEE 8th Intl
  Conference on Big Data Security on Cloud (BigDataSecurity), IEEE Intl
  Conference on High Performance and Smart Computing, (HPSC) and IEEE Intl
  Conference on Intelligent Data and Security (IDS)}}.
\newblock
\urldef\tempurl%
\url{https://doi.org/10.1109/BigDataSecurityHPSCIDS54978.2022.00042}
\showDOI{\tempurl}


\bibitem[Liang et~al\mbox{.}(2017)]%
        {Liang2017}
\bibfield{author}{\bibinfo{person}{Tengfei Liang}, \bibinfo{person}{Zhonghua
  Hu}, {and} \bibinfo{person}{Qiang Liu}.} \bibinfo{year}{2017}\natexlab{}.
\newblock \showarticletitle{Frontal Theta Activity Supports Detecting
  Mismatched Information in Visual Working Memory}.
\newblock \bibinfo{journal}{\emph{Frontiers in Psychology}}
  \bibinfo{volume}{8} (\bibinfo{date}{10} \bibinfo{year}{2017}).
\newblock
\urldef\tempurl%
\url{https://doi.org/10.3389/fpsyg.2017.01821}
\showDOI{\tempurl}


\bibitem[Lindig-León et~al\mbox{.}(2020)]%
        {LindigLeon2020}
\bibfield{author}{\bibinfo{person}{Claudia Lindig-León},
  \bibinfo{person}{Sylvain Rimbert}, {and} \bibinfo{person}{Laurent Bougrain}.}
  \bibinfo{year}{2020}\natexlab{}.
\newblock \showarticletitle{Multiclass Classification Based on Combined Motor
  Imageries}.
\newblock \bibinfo{journal}{\emph{Frontiers in Neuroscience}}
  \bibinfo{volume}{14} (\bibinfo{year}{2020}), \bibinfo{pages}{559858}.
\newblock
\urldef\tempurl%
\url{https://doi.org/10.3389/fnins.2020.559858}
\showDOI{\tempurl}
\newblock
\shownote{Published 2020 Nov 19}.


\bibitem[Lotte et~al\mbox{.}(2018)]%
        {lotte2018review}
\bibfield{author}{\bibinfo{person}{Fabien Lotte}, \bibinfo{person}{Laurent
  Bougrain}, \bibinfo{person}{Andrzej Cichocki}, \bibinfo{person}{Maureen
  Clerc}, \bibinfo{person}{Marco Congedo}, \bibinfo{person}{Alain
  Rakotomamonjy}, {and} \bibinfo{person}{Florian Yger}.}
  \bibinfo{year}{2018}\natexlab{}.
\newblock \showarticletitle{A review of classification algorithms for EEG-based
  brain--computer interfaces: a 10 year update}.
\newblock \bibinfo{journal}{\emph{Journal of neural engineering}}
  \bibinfo{volume}{15}, \bibinfo{number}{3} (\bibinfo{year}{2018}),
  \bibinfo{pages}{031005}.
\newblock


\bibitem[Mack et~al\mbox{.}(2020a)]%
        {Mack2020-cn}
\bibfield{author}{\bibinfo{person}{Michael~L Mack}, \bibinfo{person}{Alison~R
  Preston}, {and} \bibinfo{person}{Bradley~C Love}.}
  \bibinfo{year}{2020}\natexlab{a}.
\newblock \showarticletitle{Ventromedial prefrontal cortex compression during
  concept learning}.
\newblock \bibinfo{journal}{\emph{Nat. Commun.}} \bibinfo{volume}{11},
  \bibinfo{number}{1} (\bibinfo{date}{Jan.} \bibinfo{year}{2020}),
  \bibinfo{pages}{46}.
\newblock


\bibitem[Mack et~al\mbox{.}(2020b)]%
        {Mack2020}
\bibfield{author}{\bibinfo{person}{Michael~L. Mack}, \bibinfo{person}{Alison~R.
  Preston}, {and} \bibinfo{person}{Bradley~C. Love}.}
  \bibinfo{year}{2020}\natexlab{b}.
\newblock \showarticletitle{Ventromedial prefrontal cortex compression during
  concept learning}.
\newblock \bibinfo{journal}{\emph{Nature Communications}} \bibinfo{volume}{11},
  \bibinfo{number}{1} (\bibinfo{year}{2020}), \bibinfo{pages}{46}.
\newblock
\showISSN{2041-1723}
\urldef\tempurl%
\url{https://doi.org/10.1038/s41467-019-13930-8}
\showDOI{\tempurl}


\bibitem[McAuley et~al\mbox{.}(2015)]%
        {mcauley2015}
\bibfield{author}{\bibinfo{person}{Julian McAuley}, \bibinfo{person}{Rahul
  Pandey}, {and} \bibinfo{person}{Jure Leskovec}.}
  \bibinfo{year}{2015}\natexlab{}.
\newblock \showarticletitle{Inferring Networks of Substitutable and
  Complementary Products}. In \bibinfo{booktitle}{\emph{Proceedings of the 21th
  ACM SIGKDD International Conference on Knowledge Discovery and Data Mining}}
  (Sydney, NSW, Australia) \emph{(\bibinfo{series}{KDD '15})}.
  \bibinfo{publisher}{Association for Computing Machinery},
  \bibinfo{address}{New York, NY, USA}, \bibinfo{pages}{785–794}.
\newblock
\showISBNx{9781450336642}
\urldef\tempurl%
\url{https://doi.org/10.1145/2783258.2783381}
\showDOI{\tempurl}


\bibitem[Moshfeghi and Pollick(2021)]%
        {Moshfeghi21}
\bibfield{author}{\bibinfo{person}{Yashar Moshfeghi} {and}
  \bibinfo{person}{Frank Pollick}.} \bibinfo{year}{2021}\natexlab{}.
\newblock \showarticletitle{Neural Correlates of Realisation of Satisfaction in
  a Successful Search Process}.
\newblock \bibinfo{journal}{\emph{Proceedings of the Association for
  Information Science and Technology}}  \bibinfo{volume}{58}
  (\bibinfo{date}{10} \bibinfo{year}{2021}), \bibinfo{pages}{282--291}.
\newblock
\urldef\tempurl%
\url{https://doi.org/10.1002/pra2.456}
\showDOI{\tempurl}


\bibitem[Ni et~al\mbox{.}(2019)]%
        {ni-etal-2019-justifying}
\bibfield{author}{\bibinfo{person}{Jianmo Ni}, \bibinfo{person}{Jiacheng Li},
  {and} \bibinfo{person}{Julian McAuley}.} \bibinfo{year}{2019}\natexlab{}.
\newblock \showarticletitle{Justifying Recommendations using Distantly-Labeled
  Reviews and Fine-Grained Aspects}. In \bibinfo{booktitle}{\emph{Proceedings
  of the 2019 Conference on Empirical Methods in Natural Language Processing
  and the 9th International Joint Conference on Natural Language Processing
  (EMNLP-IJCNLP)}}, \bibfield{editor}{\bibinfo{person}{Kentaro Inui},
  \bibinfo{person}{Jing Jiang}, \bibinfo{person}{Vincent Ng}, {and}
  \bibinfo{person}{Xiaojun Wan}} (Eds.). \bibinfo{publisher}{Association for
  Computational Linguistics}, \bibinfo{address}{Hong Kong, China}.
\newblock
\urldef\tempurl%
\url{https://doi.org/10.18653/v1/D19-1018}
\showDOI{\tempurl}


\bibitem[Nolan et~al\mbox{.}(2010)]%
        {nolan2010faster}
\bibfield{author}{\bibinfo{person}{H. Nolan}, \bibinfo{person}{R. Whelan},
  {and} \bibinfo{person}{R.~B. Reilly}.} \bibinfo{year}{2010}\natexlab{}.
\newblock \showarticletitle{FASTER: Fully Automated Statistical Thresholding
  for EEG artifact Rejection}.
\newblock \bibinfo{journal}{\emph{Journal of Neuroscience Methods}}
  \bibinfo{volume}{192}, \bibinfo{number}{1} (\bibinfo{year}{2010}),
  \bibinfo{pages}{152--162}.
\newblock
\urldef\tempurl%
\url{https://doi.org/10.1016/j.jneumeth.2010.07.015}
\showDOI{\tempurl}


\bibitem[Oard and Kim(1998)]%
        {Oard1998ImplicitFF}
\bibfield{author}{\bibinfo{person}{Douglas~W. Oard} {and}
  \bibinfo{person}{Jinmook Kim}.} \bibinfo{year}{1998}\natexlab{}.
\newblock \showarticletitle{Implicit Feedback for Recommender Systems}.
\newblock
\urldef\tempurl%
\url{https://api.semanticscholar.org/CorpusID:8987583}
\showURL{%
\tempurl}


\bibitem[Pantazatos et~al\mbox{.}(2012)]%
        {Pantazatos2012-vo}
\bibfield{author}{\bibinfo{person}{Spiro~P Pantazatos}, \bibinfo{person}{Ted~K
  Yanagihara}, \bibinfo{person}{Xian Zhang}, \bibinfo{person}{Thomas Meitzler},
  {and} \bibinfo{person}{Joy Hirsch}.} \bibinfo{year}{2012}\natexlab{}.
\newblock \showarticletitle{Frontal-occipital connectivity during visual
  search}.
\newblock \bibinfo{journal}{\emph{Brain Connect.}} \bibinfo{volume}{2},
  \bibinfo{number}{3} (\bibinfo{date}{July} \bibinfo{year}{2012}),
  \bibinfo{pages}{164--175}.
\newblock


\bibitem[Pappalettera et~al\mbox{.}(2023)]%
        {Pappalettera2023}
\bibfield{author}{\bibinfo{person}{C. Pappalettera}, \bibinfo{person}{A.
  Cacciotti}, \bibinfo{person}{L. Nucci}, \bibinfo{person}{F. Miraglia},
  \bibinfo{person}{P.~M. Rossini}, {and} \bibinfo{person}{F. Vecchio}.}
  \bibinfo{year}{2023}\natexlab{}.
\newblock \showarticletitle{Approximate entropy analysis across
  electroencephalographic rhythmic frequency bands during physiological aging
  of human brain}.
\newblock \bibinfo{journal}{\emph{Geroscience}} \bibinfo{volume}{45},
  \bibinfo{number}{2} (\bibinfo{date}{April} \bibinfo{year}{2023}),
  \bibinfo{pages}{1131--1145}.
\newblock
\urldef\tempurl%
\url{https://doi.org/10.1007/s11357-022-00710-4}
\showDOI{\tempurl}


\bibitem[Patel and Patel(2020)]%
        {93.1}
\bibfield{author}{\bibinfo{person}{Krupa Patel} {and} \bibinfo{person}{Hiren~B.
  Patel}.} \bibinfo{year}{2020}\natexlab{}.
\newblock \showarticletitle{A state-of-the-art survey on recommendation system
  and prospective extensions}.
\newblock \bibinfo{journal}{\emph{Computers and Electronics in Agriculture}}
  \bibinfo{volume}{178} (\bibinfo{year}{2020}), \bibinfo{pages}{105779}.
\newblock
\showISSN{0168-1699}
\urldef\tempurl%
\url{https://doi.org/10.1016/j.compag.2020.105779}
\showDOI{\tempurl}


\bibitem[Peirce et~al\mbox{.}(2019)]%
        {peirce2019psychopy2}
\bibfield{author}{\bibinfo{person}{Jonathan Peirce}, \bibinfo{person}{Jeremy~R
  Gray}, \bibinfo{person}{Sol Simpson}, \bibinfo{person}{Michael MacAskill},
  \bibinfo{person}{Richard H{\"o}chenberger}, \bibinfo{person}{Hiroyuki Sogo},
  \bibinfo{person}{Erik Kastman}, {and} \bibinfo{person}{Jonas~Kristoffer
  Lindel{\o}v}.} \bibinfo{year}{2019}\natexlab{}.
\newblock \showarticletitle{PsychoPy2: Experiments in behavior made easy}.
\newblock \bibinfo{journal}{\emph{Behavior research methods}}
  \bibinfo{volume}{51} (\bibinfo{year}{2019}), \bibinfo{pages}{195--203}.
\newblock


\bibitem[Peng et~al\mbox{.}(1994)]%
        {PhysRevE.49.1685}
\bibfield{author}{\bibinfo{person}{C.-K. Peng}, \bibinfo{person}{S.~V.
  Buldyrev}, \bibinfo{person}{S. Havlin}, \bibinfo{person}{M. Simons},
  \bibinfo{person}{H.~E. Stanley}, {and} \bibinfo{person}{A.~L. Goldberger}.}
  \bibinfo{year}{1994}\natexlab{}.
\newblock \showarticletitle{Mosaic organization of DNA nucleotides}.
\newblock \bibinfo{journal}{\emph{Phys. Rev. E}}  \bibinfo{volume}{49}
  (\bibinfo{date}{Feb} \bibinfo{year}{1994}), \bibinfo{pages}{1685--1689}.
\newblock
Issue 2.
\urldef\tempurl%
\url{https://doi.org/10.1103/PhysRevE.49.1685}
\showDOI{\tempurl}


\bibitem[Pereda-Ba{\~n}os et~al\mbox{.}(2015)]%
        {pereda2015human}
\bibfield{author}{\bibinfo{person}{Alexandre Pereda-Ba{\~n}os},
  \bibinfo{person}{Ioannis Arapakis}, {and} \bibinfo{person}{Miguel
  Barreda-{\'A}ngeles}.} \bibinfo{year}{2015}\natexlab{}.
\newblock \showarticletitle{On Human Information Processing in Information
  Retrieval (Position Paper)}. In \bibinfo{booktitle}{\emph{SIGIR Workshop on
  Neuro-Physiological Methods in IR Research (NeuroIR)}}. ACM.
\newblock


\bibitem[Pinkosova et~al\mbox{.}(2023)]%
        {Pinkosova2023}
\bibfield{author}{\bibinfo{person}{Zuzana Pinkosova},
  \bibinfo{person}{William~J. McGeown}, {and} \bibinfo{person}{Yashar
  Moshfeghi}.} \bibinfo{year}{2023}\natexlab{}.
\newblock \showarticletitle{Revisiting Neurological Aspects of Relevance: An
  EEG Study}. In \bibinfo{booktitle}{\emph{Machine Learning, Optimization, and
  Data Science}}. \bibinfo{publisher}{Springer Nature Switzerland}.
\newblock
\showISBNx{978-3-031-25891-6}


\bibitem[Qu and Nobuhara(2024)]%
        {10545583}
\bibfield{author}{\bibinfo{person}{Yuanpeng Qu} {and} \bibinfo{person}{Hajime
  Nobuhara}.} \bibinfo{year}{2024}\natexlab{}.
\newblock \showarticletitle{Generating Explanations for Explainable
  Recommendations Using Filter-Enhanced Time-Series Information}.
\newblock \bibinfo{journal}{\emph{IEEE Access}}  \bibinfo{volume}{12}
  (\bibinfo{year}{2024}), \bibinfo{pages}{78480--78495}.
\newblock
\urldef\tempurl%
\url{https://doi.org/10.1109/ACCESS.2024.3408252}
\showDOI{\tempurl}


\bibitem[Reddy et~al\mbox{.}(2022)]%
        {Reddy2022}
\bibfield{author}{\bibinfo{person}{Chandan~K. Reddy}, \bibinfo{person}{Lluís
  Màrquez}, \bibinfo{person}{Fran Valero}, \bibinfo{person}{Nikhil Rao},
  \bibinfo{person}{Hugo Zaragoza}, \bibinfo{person}{Sambaran Bandyopadhyay},
  \bibinfo{person}{Arnab Biswas}, \bibinfo{person}{Anlu Xing}, {and}
  \bibinfo{person}{Karthik Subbian}.} \bibinfo{year}{2022}\natexlab{}.
\newblock \bibinfo{title}{Shopping Queries Dataset: A Large-Scale ESCI
  Benchmark for Improving Product Search}.
\newblock
\newblock
\showeprint[arxiv]{2206.06588}~[cs.IR]
\urldef\tempurl%
\url{https://arxiv.org/abs/2206.06588}
\showURL{%
\tempurl}


\bibitem[Ruotsalo et~al\mbox{.}(2023)]%
        {tukka2023}
\bibfield{author}{\bibinfo{person}{Tuukka Ruotsalo}, \bibinfo{person}{Kalle
  M\"{a}kel\"{a}}, \bibinfo{person}{Michiel~M. Spap\'{e}}, {and}
  \bibinfo{person}{Luis~A. Leiva}.} \bibinfo{year}{2023}\natexlab{}.
\newblock \showarticletitle{Affective Relevance: Inferring Emotional Responses
  via fNIRS Neuroimaging}. In \bibinfo{booktitle}{\emph{Proceedings of the 46th
  International ACM SIGIR Conference on Research and Development in Information
  Retrieval}} \emph{(\bibinfo{series}{SIGIR '23})}.
  \bibinfo{publisher}{Association for Computing Machinery},
  \bibinfo{address}{New York, NY, USA}.
\newblock
\showISBNx{9781450394086}
\urldef\tempurl%
\url{https://doi.org/10.1145/3539618.3591946}
\showDOI{\tempurl}


\bibitem[Schwartz(2004)]%
        {schwartz2004paradox}
\bibfield{author}{\bibinfo{person}{Barry Schwartz}.}
  \bibinfo{year}{2004}\natexlab{}.
\newblock \showarticletitle{The paradox of choice: Why less is more}.
\newblock \bibinfo{journal}{\emph{New York: Ecco}} (\bibinfo{year}{2004}).
\newblock


\bibitem[Srivastava et~al\mbox{.}(2013)]%
        {srivastava2013support}
\bibfield{author}{\bibinfo{person}{Tanu Srivastava}, \bibinfo{person}{Shilpi
  Singh}, {and} \bibinfo{person}{Sanjeev Bhardwaj}.}
  \bibinfo{year}{2013}\natexlab{}.
\newblock \showarticletitle{Support vector machine technique for EEG signals}.
\newblock \bibinfo{journal}{\emph{International Journal of Engineering Research
  and Applications}} \bibinfo{volume}{3}, \bibinfo{number}{1}
  (\bibinfo{year}{2013}), \bibinfo{pages}{533--538}.
\newblock


\bibitem[Stamenkovic et~al\mbox{.}(2021)]%
        {94}
\bibfield{author}{\bibinfo{person}{Dusan Stamenkovic},
  \bibinfo{person}{Alexandros Karatzoglou}, \bibinfo{person}{Ioannis Arapakis},
  \bibinfo{person}{Xin Xin}, {and} \bibinfo{person}{Kleomenis Katevas}.}
  \bibinfo{year}{2021}\natexlab{}.
\newblock \showarticletitle{Choosing the Best of Both Worlds: Diverse and Novel
  Recommendations through Multi-Objective Reinforcement Learning}.
\newblock \bibinfo{journal}{\emph{CoRR}}  \bibinfo{volume}{abs/2110.15097}
  (\bibinfo{year}{2021}).
\newblock
\urldef\tempurl%
\url{https://arxiv.org/abs/2110.15097}
\showURL{%
\tempurl}


\bibitem[Subasi et~al\mbox{.}(2017)]%
        {subasi2017classification}
\bibfield{author}{\bibinfo{person}{Abdulhamit Subasi}, \bibinfo{person}{Ahmet
  Alkaya}, \bibinfo{person}{Rahime Keskin}, {and} \bibinfo{person}{Etem
  Koklukaya}.} \bibinfo{year}{2017}\natexlab{}.
\newblock \showarticletitle{Classification of EEG signals based on pattern
  recognition approach}.
\newblock \bibinfo{journal}{\emph{International Journal of Imaging Systems and
  Technology}} \bibinfo{volume}{27}, \bibinfo{number}{3}
  (\bibinfo{year}{2017}), \bibinfo{pages}{213--221}.
\newblock


\bibitem[Sutton et~al\mbox{.}(1965)]%
        {sutton1965evoked}
\bibfield{author}{\bibinfo{person}{Samuel Sutton}, \bibinfo{person}{Margery
  Braren}, \bibinfo{person}{Joseph Zubin}, {and} \bibinfo{person}{ER John}.}
  \bibinfo{year}{1965}\natexlab{}.
\newblock \showarticletitle{Evoked-potential correlates of stimulus
  uncertainty}.
\newblock \bibinfo{journal}{\emph{Science}} \bibinfo{volume}{150},
  \bibinfo{number}{3700} (\bibinfo{year}{1965}), \bibinfo{pages}{1187--1188}.
\newblock


\bibitem[Tadson et~al\mbox{.}(2023)]%
        {14}
\bibfield{author}{\bibinfo{person}{Bella Tadson}, \bibinfo{person}{Jared
  Boasen}, \bibinfo{person}{Fran{\c{c}}ois Courtemanche},
  \bibinfo{person}{No{\'e}mie Beauchemin}, \bibinfo{person}{Alexander-John
  Karran}, \bibinfo{person}{Pierre-Majorique L{\'e}ger}, {and}
  \bibinfo{person}{Sylvain S{\'e}n{\'e}cal}.} \bibinfo{year}{2023}\natexlab{}.
\newblock \showarticletitle{Neuro-Adaptive Interface System to Evaluate Product
  Recommendations in the Context of E-Commerce}. In
  \bibinfo{booktitle}{\emph{Design Science Research for a New Society: Society
  5.0}}, \bibfield{editor}{\bibinfo{person}{Aurona Gerber} {and}
  \bibinfo{person}{Richard Baskerville}} (Eds.).
\newblock
\showISBNx{978-3-031-32808-4}


\bibitem[Tang and Li(2024)]%
        {Tang2024}
\bibfield{author}{\bibinfo{person}{Shaohua Tang} {and} \bibinfo{person}{Zheng
  Li}.} \bibinfo{year}{2024}\natexlab{}.
\newblock \showarticletitle{EEG complexity measures for detecting mind
  wandering during video-based learning}.
\newblock \bibinfo{journal}{\emph{Scientific Reports}} \bibinfo{volume}{14},
  \bibinfo{number}{1} (\bibinfo{date}{8 4} \bibinfo{year}{2024}),
  \bibinfo{pages}{8209}.
\newblock
\showISSN{2045-2322}
\urldef\tempurl%
\url{https://doi.org/10.1038/s41598-024-58889-9}
\showDOI{\tempurl}


\bibitem[Thibodeau et~al\mbox{.}(2006)]%
        {Thibodeau2006}
\bibfield{author}{\bibinfo{person}{Ryan Thibodeau}, \bibinfo{person}{Randall~S.
  Jorgensen}, {and} \bibinfo{person}{Sanghag Kim}.}
  \bibinfo{year}{2006}\natexlab{}.
\newblock \showarticletitle{Depression, anxiety, and resting frontal EEG
  asymmetry: A meta-analytic review}.
\newblock \bibinfo{journal}{\emph{Journal of Abnormal Psychology}}
  \bibinfo{volume}{115}, \bibinfo{number}{4} (\bibinfo{year}{2006}),
  \bibinfo{pages}{715--729}.
\newblock
\urldef\tempurl%
\url{https://doi.org/10.1037/0021-843X.115.4.715}
\showDOI{\tempurl}


\bibitem[Wallace et~al\mbox{.}(2022)]%
        {Wallace2022}
\bibfield{author}{\bibinfo{person}{Shaun Wallace}, \bibinfo{person}{Zoya
  Bylinskii}, \bibinfo{person}{Jonathan Dobres}, \bibinfo{person}{Bernard
  Kerr}, \bibinfo{person}{Sam Berlow}, \bibinfo{person}{Rick Treitman},
  \bibinfo{person}{Nirmal Kumawat}, \bibinfo{person}{Kathleen Arpin},
  \bibinfo{person}{Dave~B. Miller}, \bibinfo{person}{Jeff Huang}, {and}
  \bibinfo{person}{Ben~D. Sawyer}.} \bibinfo{year}{2022}\natexlab{}.
\newblock \showarticletitle{Towards Individuated Reading Experiences: Different
  Fonts Increase Reading Speed for Different Individuals}.
\newblock \bibinfo{journal}{\emph{ACM Trans. Comput.-Hum. Interact.}}
  \bibinfo{volume}{29}, \bibinfo{number}{4}, Article \bibinfo{articleno}{38}
  (\bibinfo{date}{March} \bibinfo{year}{2022}), \bibinfo{numpages}{56}~pages.
\newblock
\showISSN{1073-0516}
\urldef\tempurl%
\url{https://doi.org/10.1145/3502222}
\showDOI{\tempurl}


\bibitem[Wei et~al\mbox{.}(2016)]%
        {wei2016classification}
\bibfield{author}{\bibinfo{person}{Chun-Shu Wei}, \bibinfo{person}{Yuan-Pin
  Lin}, \bibinfo{person}{Yu-Te Wang}, \bibinfo{person}{Tzyy-Ping Jung},
  \bibinfo{person}{Nima Bigdely-Shamlo}, {and} \bibinfo{person}{Chin-Teng
  Lin}.} \bibinfo{year}{2016}\natexlab{}.
\newblock \showarticletitle{Classification of motor imagery EEG signals with
  support vector machines and particle swarm optimization}.
\newblock \bibinfo{journal}{\emph{Computational and mathematical methods in
  medicine}}  \bibinfo{volume}{2016} (\bibinfo{year}{2016}).
\newblock


\bibitem[White et~al\mbox{.}(2002)]%
        {white2002comparing}
\bibfield{author}{\bibinfo{person}{Ryen~W White}, \bibinfo{person}{Joemon~M
  Jose}, \bibinfo{person}{Ian Ruthven}, \bibinfo{person}{EM Voorhees}, {and}
  \bibinfo{person}{DK Harman}.} \bibinfo{year}{2002}\natexlab{}.
\newblock \showarticletitle{Comparing explicit and implicit feedback techniques
  for web retrieval: Trec-10 interactive track report}. In
  \bibinfo{booktitle}{\emph{Proceedings of the tenth text retrieval conference
  (TREC-10)}}.
\newblock


\bibitem[Yan et~al\mbox{.}(2022)]%
        {Yan2022}
\bibfield{author}{\bibinfo{person}{An Yan}, \bibinfo{person}{Chaosheng Dong},
  \bibinfo{person}{Yan Gao}, \bibinfo{person}{Jinmiao Fu},
  \bibinfo{person}{Tong Zhao}, \bibinfo{person}{Yi Sun}, {and}
  \bibinfo{person}{Julian McAuley}.} \bibinfo{year}{2022}\natexlab{}.
\newblock \showarticletitle{Personalized complementary product recommendation}.
\newblock  (\bibinfo{year}{2022}).
\newblock
\urldef\tempurl%
\url{https://www.amazon.science/publications/personalized-complementary-product-recommendation}
\showURL{%
\tempurl}


\bibitem[Yang et~al\mbox{.}(2025)]%
        {Yang2025}
\bibfield{author}{\bibinfo{person}{Banghua Yang}, \bibinfo{person}{Fenqi Rong},
  \bibinfo{person}{Yunlong Xie}, \bibinfo{person}{Du Li},
  \bibinfo{person}{Jiayang Zhang}, \bibinfo{person}{Fu Li},
  \bibinfo{person}{Guangming Shi}, {and} \bibinfo{person}{Xiaorong Gao}.}
  \bibinfo{year}{2025}\natexlab{}.
\newblock \showarticletitle{A multi-day and high-quality EEG dataset for motor
  imagery brain-computer interface}.
\newblock \bibinfo{journal}{\emph{Scientific Data}} \bibinfo{volume}{12},
  \bibinfo{number}{1} (\bibinfo{year}{2025}), \bibinfo{pages}{488}.
\newblock
\showISSN{2052-4463}
\urldef\tempurl%
\url{https://doi.org/10.1038/s41597-025-04826-y}
\showDOI{\tempurl}


\bibitem[Ye et~al\mbox{.}(2024)]%
        {ye2024brain}
\bibfield{author}{\bibinfo{person}{Ziyi Ye}, \bibinfo{person}{Qingyao Ai},
  {and} \bibinfo{person}{Yiqun Liu}.} \bibinfo{year}{2024}\natexlab{}.
\newblock \showarticletitle{Brain-Computer Interface Meets Information
  Retrieval: Perspective on Next-generation Information System}. In
  \bibinfo{booktitle}{\emph{Proceedings of the 1st International Workshop on
  Brain-Computer Interfaces (BCI) for Multimedia Understanding}}.
  \bibinfo{pages}{61--65}.
\newblock


\bibitem[Ye et~al\mbox{.}(2022)]%
        {ye2022don}
\bibfield{author}{\bibinfo{person}{Ziyi Ye}, \bibinfo{person}{Xiaohui Xie},
  \bibinfo{person}{Yiqun Liu}, \bibinfo{person}{Zhihong Wang},
  \bibinfo{person}{Xuancheng Li}, \bibinfo{person}{Jiaji Li},
  \bibinfo{person}{Xuesong Chen}, \bibinfo{person}{Min Zhang}, {and}
  \bibinfo{person}{Shaoping Ma}.} \bibinfo{year}{2022}\natexlab{}.
\newblock \showarticletitle{Why Don't You Click: Understanding Non-Click
  Results in Web Search with Brain Signals}. In
  \bibinfo{booktitle}{\emph{Proceedings of the 45th International ACM SIGIR
  Conference on Research and Development in Information Retrieval}}.
  \bibinfo{pages}{633--645}.
\newblock


\bibitem[Yuan et~al\mbox{.}(2019)]%
        {yuan2020}
\bibfield{author}{\bibinfo{person}{Fajie Yuan}, \bibinfo{person}{Alexandros
  Karatzoglou}, \bibinfo{person}{Ioannis Arapakis}, \bibinfo{person}{Joemon~M
  Jose}, {and} \bibinfo{person}{Xiangnan He}.} \bibinfo{year}{2019}\natexlab{}.
\newblock \showarticletitle{A simple convolutional generative network for next
  item recommendation}. In \bibinfo{booktitle}{\emph{Proceedings of the Twelfth
  ACM International Conference on Web Search and Data Mining}}.
  \bibinfo{pages}{582--590}.
\newblock


\bibitem[Zhang et~al\mbox{.}(2024)]%
        {zhang2024eeggan}
\bibfield{author}{\bibinfo{person}{Yingchun Zhang}, \bibinfo{person}{Xin Wang},
  \bibinfo{person}{Yiwen Jiang}, {and} \bibinfo{person}{Xiaowei Xu}.}
  \bibinfo{year}{2024}\natexlab{}.
\newblock \showarticletitle{EEGGAN-Net: enhancing EEG signal classification
  through data augmentation framework}.
\newblock \bibinfo{journal}{\emph{Frontiers in Human Neuroscience}}
  \bibinfo{volume}{18} (\bibinfo{year}{2024}), \bibinfo{pages}{1430086}.
\newblock
\urldef\tempurl%
\url{https://doi.org/10.3389/fnhum.2024.1430086}
\showDOI{\tempurl}


\bibitem[Zhang et~al\mbox{.}(2020)]%
        {zhang2020eeg}
\bibfield{author}{\bibinfo{person}{Yong Zhang}, \bibinfo{person}{Yong Wang},
  \bibinfo{person}{Jianting Jin}, {and} \bibinfo{person}{Xingyu Wang}.}
  \bibinfo{year}{2020}\natexlab{}.
\newblock \showarticletitle{EEG signal classification based on SVM with
  improved squirrel search algorithm}.
\newblock \bibinfo{journal}{\emph{Computational intelligence and neuroscience}}
   \bibinfo{volume}{2020} (\bibinfo{year}{2020}).
\newblock


\bibitem[Zúñiga et~al\mbox{.}(2024)]%
        {Zuniga2024}
\bibfield{author}{\bibinfo{person}{M.~A. Zúñiga}, \bibinfo{person}{Á.
  Acero-González}, \bibinfo{person}{J.~C. Restrepo-Castro},
  \bibinfo{person}{M.~Á. Uribe-Laverde}, \bibinfo{person}{D.~A. Botero-Rosas},
  \bibinfo{person}{B.~I. Ferreras}, \bibinfo{person}{M.~C. Molina-Borda}, {and}
  \bibinfo{person}{M.~P. Villa-Reyes}.} \bibinfo{year}{2024}\natexlab{}.
\newblock \showarticletitle{Is EEG Entropy a Useful Measure for Alzheimer's
  Disease?}
\newblock \bibinfo{journal}{\emph{Actas Esp Psiquiatr}} (\bibinfo{year}{2024}).
\newblock
\urldef\tempurl%
\url{https://doi.org/10.62641/aep.v52i3.1632}
\showDOI{\tempurl}


\end{thebibliography}
